\documentstyle[12pt]{article}
\newcommand{\resection}[1]{\setcounter{equation}{0}\section{#1}}

\renewcommand{\thefootnote}{\fnsymbol{footnote}}
\newtheorem{pro}{Proposition}
\newtheorem{lm}{Lemma}
\newtheorem{th}{Theorem}
\newcommand{\beq}{\begin{equation}}
\newcommand{\eeq}{\end{equation}}
\newcommand{\beqn}{\begin{eqnarray}}
\newcommand{\eeqn}{\end{eqnarray}}
\newcommand{\la}[1]{\label{#1}}

\def\skphs{SKP hierarchies}
\def\spdo{S$\Psi$DO}
\def\spdos{S$\Psi$DO's}
\def\sG{super-Grassmannian}
\def\H{{\cal H}}
\def\ala{{\cal A}}
\def\E{{\cal E}}
\def\D{{\cal D}}
\def\G{{Gr_{0}}}
\def\sgrass{\Gamma_0}
\def\bZ{{\bf Z}}
\def\bZzp{\bZ_{\geq}}
\def\bZm{\bZ_{<}}
\def\tilrho{\tilde \rho}
\def\sop{1+s_{-1}D^{-1}+s_{-2}D^{-2}+\cdots}
\def\op[#1,(#2;#3)]{\sum_{#2}{#1}_{#3}\,D^{#3} }
\def\complnum{{\bf C}}
\def\ring{\complnum [[ x,\xi]]}

\def\der-#1{\partial_{#1}}
\def\pder[#1]-#2{{\der-#2}^{-#1}}
\def\pd[#1]{D^{-#1}}
\def\gauss(#1,#2){\left[\frac{#1}{#2}\right]}
\def\xximat{x\Lambda + \xi\Gamma}

\def\xxi{xz + \xi\theta}

\def\oovpi{\frac{1}{2\pi i}}
\def\matZZR{Mat(\bZ\times\bZ,\>R)}
\def\matZZA{Mat(\bZ\times\bZ,\>\ala)}
\def\evol[#1]{U_{#1}}
\def\hamil[#1]{H_{#1}}
\def\pt-#1{t^+_{#1}}
\def\mt-#1{t^-_{#1}}
\def\pmt-#1{t^\pm_{#1}}
\def\mpt-#1{t^\mp_{#1}}
\def\podd-#1{{2#1+1}}
\def\modd-#1{{2#1-1}}
\def\leinf{{}^\leq}
\def\wleinf{{}^{\leq\leq}}
\def\ltinf{{}^<}
\def\wltinf{{}^{<<}}
\def\geinf{{}^\geq}
\def\wgeinf{{}^{\geq\geq}}
\def\gtinf{{}^>}

\def\leltinf{{}^{\leq <}}

\def\geltinf{{}^{\geq <}}

\def\ltgeinf{{}^{<\geq}}
\def\pJ{\frac{1+J}{2}}
\def\mJ{\frac{1-J}{2}}
\def\xxitz{(x,\xi,t;\,\zeta)}
\def\xxipmtz{(x,\xi,\pmt-{}\,;\zeta)}
\def\zexxipmt{(\zeta;\,x,\xi,\pmt-{})}
\def\zxxipmt{(z;\,x,\xi,\pmt-{})}

\def\zpmt{(z;\,\pmt-{})}

\def\wvf{w\xxitz}
\def\supwvf-#1{w_{{}_{#1}}\xxitz}
\def\waveM{w_{{}_M}\xxipmtz}
\def\dwaveM{{\widetilde w}_{{}_M}\xxipmtz}

\def\tauM{\tau_{{}_M}}
\def\dwvf{{\widetilde w}\xxitz}
\def\facA{A\xxitz}
\def\dfacA{{\widetilde A}\xxitz}
\def\facAM{A_M\xxipmtz}
\def\dfacAM{{\widetilde A_M}\xxipmtz}
\def\pwpte-#1{\pt-{e}(#1)}
\def\pwmte-#1{\mt-{e}(#1)}
\def\pwpmte-#1{\pmt-{e}(#1)}
\def\pwptd-#1{\pt-{o}(#1)}
\def\pwmtd-#1{\mt-{o}(#1)}
\def\pwpmtd-#1{\pmt-{o}(#1)}
\def\pwmpte-#1{\mpt-{e}(#1)}
\def\pwte-#1{t_{e}(#1)}
\def\pwbarte-#1{{\bar t}_{e}(#1)}
\def\pwtd-#1{t_{o}(#1)}
\def\pwbartd-#1{{\bar t}_{o}(#1)}
\def\pwxipt-#1{\pt-{\xi}(#1)}
\def\pwximt-#1{\mt-{\xi}(#1)}
\def\pwxipmt-#1{\pmt-{\xi}(#1)}
\def\propwtd-#1{\pt-{o}\pwmtd-#1}
\def\dualframe{{\widetilde F_0}}
\def\phalf-#1{{#1+\frac12}}
\def\mhalf-#1{{#1-\frac12}}
\def\bep-#1{\beta_{\phalf-{#1}}}
\def\bem-#1{\beta_{\phalf-{#1}}}
\def\gamp-#1{\gamma_{\phalf-#1}}
\def\gamm-#1{\gamma_{\mhalf-#1}}
\def\alBC{{\bf A}_{BC}}
\def\albega{{\bf A}_{\beta\gamma}}
\def\albc{{\bf A}_{bc}}
\def\aljpsi{{\bf A}_{j\psi}}
\def\spaceH{{\bf H}}
\def\pj-#1{j^+_{#1}}
\def\mj-#1{j^-_{#1}}
\def\pmj-#1{j^\pm_{#1}}
\def\tj-#1{j^t_{#1}}
\def\ppsi-#1{\psi^+_{#1}}
\def\mpsi-#1{\psi^-_{#1}}
\def\pmpsi-#1{\psi^\pm_{#1}}
\def\pmpsip-#1{\psi^\pm_{\phalf-{#1}}}
\def\ppsip-#1{\psi^+_{\phalf-{#1}}}
\def\ppsim-#1{\psi^+_{\mhalf-{#1}}}
\def\mpsip-#1{\psi^-_{\phalf-{#1}}}
\def\mpsim-#1{\psi^-_{\mhalf-{#1}}}
\def\gl11{{\widehat {gl1|1}}}
\def\ketbc-#1{{#1}\rangle\!_{{}_{bc}}}
\def\ketbega-#1{{#1}\rangle\!_{{}_{\beta\gamma}}}
\def\brabc-#1{{}_{{}_{bc}}\!\langle {#1}}
\def\brabega-#1{{}_{{}_{\beta\gamma}}\!\langle{#1}}
\def\ketcurr-#1{{#1}\rangle}
\def\bracurr-#1{\langle {#1}}
\def\ketpq[#1,#2]{(#1,#2)\rangle}
\def\brapq[#1,#2]{\langle(#1,#2)}
\def\F{{\cal F}}
\def\tilcapc-#1{{\widetilde C_{#1}}}
\def\tilc-#1{{\tilde c}_{#1}}
\def\tilgamp-#1{{\tilde \gamma}_{\phalf-{#1}}}
\def\tilgam-#1{{\tilde \gamma}_{#1}}
\def\liea{a_{\infty|\infty}}
\def\lieabar{{\bar a}_{\infty|\infty}}
\def\opevol{{\widehat \Phi}}
\def\opxxi{x\tj-1 +\xi\sigma}
\def\opppsi{\psi^+[t^+_o]}
\def\opmpsi{\psi^-[t^-_o]}
\def\oppj{j^+[t^+_e]}
\def\opmj{j^-[t^-_e]}
\def\optj{j^t[t_e]}
\def\oppmpsi{\psi^\pm[t^\pm_o]}
\def\oppmj{j^\pm[t^\pm_e]}
\def\opO-#1{{\widehat O}({#1})}
\def\pphi-#1{\phi^+_{#1}}
\def\mphi-#1{\phi^-_{#1}}
\def\pmphi-#1{\phi^\pm_{#1}}
\def\mpphi-#1{\phi^\mp_{#1}}
\def\phiani#1{\phi^{#1}_{\gtinf}}
\def\phicri#1{\phi^{#1}_{\ltinf}}
\def\psiani#1{\psi^{#1}_{\gtinf}}

\def\trsub[#1,#2]{{}_{{}_{[#2]_#1}}\!\!}
\def\opT[#1,#2]{{\cal T}(#1;\,#2)}
\def\pwlog-#1{\sum_{n=1}\frac{-1}{n}\,{#1}^{-n}}
\def\pwslog-#1{\sum\frac{-1}{n}\,{#1}^{-n}}
\def\opd[#1,#2]{D^{#1}_{#2}}
\def\opdchk[#1,#2]{{\check D}^{#1}_{#2}}

\begin{document}
\begin{titlepage}
\nopagebreak
\begin{flushright}
YITP/U-95-23\hfill\\
June 1995\hfill
\end{flushright}

\renewcommand{\thefootnote}{\fnsymbol{footnote}}
\vfill
\begin{center}
{\Large Grassmannian Approach to Super-KP Hierarchies}\\
\vskip 20mm

{\large Michiaki TAKAMA}
\vskip 10mm
{\it Uji Research Center, Yukawa Institute for Theoretical Physics \\
  Kyoto University, Uji 611, Japan}
\end{center}
\vskip 20mm
\begin{abstract}
We present a theory of 'maximal' super-KP(SKP) hierarchy whose flows
are maximally extended to include all those of known SKP hierarchies,
including, for example, the MRSKP hierarchy of Manin and Radul and the
Jacobian SKP(JSKP) introduced by Mulase and Rabin.  It is shown that
SKP hierarchies has a natural field theoretic description in terms of
the B-C system, in analogous way as the ordinary KP hierarchy.  For
this SKP hierarchy, we construct the vertex operators by using Kac-van
de Leur superbosonization. The vertex operators act on the
\(\tau\)-function and then produce the wave function and the dual wave
function of the hierarchy. Thereby we achieve the description of the
'maximal' SKP hierarchy in terms of the \(\tau\)-function, which
seemed to be lacking till now.  Mutual relations among the SKP
hierarchies are clarified.  The MRSKP and the JSKP hierarchies are
obtained as special cases when the time variables are appropriately
restricted.
\end{abstract}
\vfill
\end{titlepage}

\resection{Introduction}
Nearly a decade has passed since Manin and Radul introduced a
supersymmetric extension of the KP hierarchy (SKP),(to be denoted
as MRSKP)\cite{man-rad}. In the meanwhile, several variations of
SKP hierarchies were born, Jacobian SKP(JSKP) hierarchy by Mulase and
Rabin\cite{mula}\cite{rabin}, SKP\({}_2\) hierarchy by
Figueroa-O'Farrill et al.\cite{figueroa}, etc.
A lot of works have been made on;
\begin{itemize}
\item
integrability and unique solvability of initial
value problems, formulating in terms of flows
on a {\sG}\cite{mula2}\cite{ike-yama},
\item
geometrical interpretation by extending the one of the ordinary KP
hierarchy in which the orbits of the flows are canonically isomorphic
to the Jacobian varieties of an algebraic curve\cite{mula}, (In this
respect JSKP hierarchy succeeds to the ordinary one, as its flows
characterize the Jacobian variety of an algebraic
\(1|1\) super curve.)
\item
construction of algebro-geometric
solutions\cite{radul}\cite{pakuliak}\cite{rabin2},
\item
reduction to the supersymmetric KdV(SKdV) hierarchy and its
Hamiltonian structure as a completely integrable
system\cite{reduction}.
\end{itemize}
Recent developments in understanding non-critical strings and the
two-dimensional quantum gravity (coupled with various conformal
fields) reveal surprising link between these theories and integrable
hierarchies of the KP type\cite{2d-grav}. It is natural to ask
whether these results can be extended to the supersymmetric ground.
A naive but not well founded speculation is that some supersymmetric
hierarchies might underlie the two-dimensional supergravities.
Though there are some proposals\cite{2d-supgrav}, we
have not yet a definite answer.  These subjects add to our interest
for the SKP hierarchies. In this respect, continued from the above
list,
\begin{itemize}
\item
additional symmetries and the relation with super\(W\)-algebras
\end{itemize}
investigated by several authors\cite{addsym}.  However with these
studies, it seems that some important elements are still missing in
the theory of the SKP hierarchies.  Indeed the most conspicuously
lacking is the concept of the
\(\tau\)-function which is a vital element of the theory and this is
related to the non-uniqueness of the SKP hierarchies.
\\ \indent
Let us recall the Grassmannian approach to the KP hierarchy and the
notion of the \(\tau\)-function\cite{djkm}.
The KP hierarchy was formulated as the eigen value preserving
deformations of a first order pseudo-differential operator
(i.e., Lax operator)
\(L(t_j)=
\der-x+u_{-1}(t_j)\der-x^{-1}+u_{-2}(t_j)\der-x^{-2}+\cdots\):
\begin{eqnarray*}
L(t_j)\;w(x,t_j;z)&=&z\,w(x,t_j;z),\nonumber\\
\der-{t_n}L(t_j)&=&(L(t_j)^n)_+L(t_j),\quad\hbox{with}\>\>t_1=x.
\nonumber\\
\end{eqnarray*}
Without loss of generality \(L\) is expressed as a dressed operator of
\(\der-x\) by a wave operator
\(S=1+s_{-1}\der-x^{-1}+s_{-2}\der-x^{-2}+\cdots\), i.e.
\(L=S\der-x S^{-1}\).
Then the wave function \(w\) is given by the formula
\(w(x,t_j;z)=S(t_j)\exp{\sum_{n=1} t_n z^n}\).
The time evolution of \(L\) is then converted to that of \(S\)
which is governed by the Sato equations:
\[
\der-{t_n}S=-(S\der-x^nS^{-1})_- S,\qquad n=1,2,\cdots.
\]
The Grassmannian nature of the KP hierarchy arises from the Sato
correspondence which says there is a one to one correspondence between
the space of the wave operators and a certain infinite dimensional
Grassmannian. As a result the KP hierarchy is interpreted as a dynamical
system on this Grassmannian, which is also considered as the solution
space of the KP hierarchy, by viewing it as the set of initial data of
the time evolution. Then \(GL(\infty)\) arises as a hidden symmetry,
that is the transformation group of this solution space.  These things
look apparent when one takes the free fermion description of the KP
hierarchy, in which the Grassmannian is projectively embedded into the
fermion Fock space and the KP flows are generated by the positive
modes of the current operator \(j_n\). Given an initial point
represented by a state \(|G\rangle\) (which sits on the \(GL(\infty)\)
orbit of the vacuum \(|0\rangle\)), an orbit of the flow is
represented by
\(\exp\sum_{n=1} t_n j_n\,|G\rangle\). The contraction with the
vacuum state gives the \(\tau\)-function
\(\tau(t_j)=\langle 0|\exp\sum_{n=1} t_n j_n\,|G\rangle\). Then
\(\tau\) maps \(|G\rangle\) to a polynomial in the time
variables, which is just the boson-fermion correspondence.  All
the information about the state \(|G\rangle\) is encoded into the
function \(\tau(t_j)\).  The locus of the Grassmannian in the fermion
Fock space is characterized by the bilinear identity
\[
\oint dz\;\psi(z)|G\rangle\otimes \psi^\ast(z)|G\rangle=0,
\]
which turns out to be the Hirota's bilinear equations for the
\(\tau\)-function. It is a great insight of Sato that reveals
the Hirota's bilinear equations to be nothing but the Pl\"uker
relations. The wave function is then identified as
\begin{eqnarray*}
w(x,t_j;z)&=&
\frac{1}{\tau(t_j)}\;
\langle 0|\psi^\ast_{\frac12}\;e^{\sum_{n=1} t_n j_n}\,\psi(z)|G\rangle
\nonumber\\
&=&\frac{1}{\tau(t_j)}\;
e^{\sum t_nz^n}\,e^{-\sum\frac{z^{-n}}{n}\der-{t_n}}\;\tau(t_j),
\nonumber\\
\end{eqnarray*}
where the vertex operator representation of the fermion field
\(\psi(z)\) is used. In this way a solution for the original
wave operator \(S(t_j)\) is obtained from the \(\tau\)-function
passing through the wave function.
\\ \indent
Turning to the SKP hierarchies, we can proceed in a parallel way.  The
SKP hierarchies are interpreted as the dynamical systems on the
{\sG}\cite{ike-yama}.  Just like the KP hierarchy allows the free
fermion description, they can be fairly well formulated in a field
theoretic terms employing the so called the B-C system, which is given
by a tensor product of the first order system of free fermion
(i.e. b-c system) and the first order system of free boson
(i.e. \(\beta\)-\(\gamma\) system).  Although we have not been able to
find the literatures which state this point explicitly by showing the
expression for the wave function\footnote{ In ref.\cite{awada}, the
SKP hierarchies are studied in terms of the B-C system, however the
description there is not fully developed.}  it seems rather evident if
one takes into account the following facts:
\begin{itemize}
\item Infinite dimensional {\sG} is realized in the Fock space
of the B-C system as the \(GL(\infty|\infty)\) orbit of the
vacuum state\cite{bergv}.
\item It is possible to characterize, through the bilinear identity, the
wave function and its dual for the SKP hierarchies\cite{ike-yama}.
\item The \(GL(\infty|\infty)\) orbit of the vacuum state of the B-C system
satisfies the bilinear identity\cite{kac-vdleur}
\[
\oint dzd\theta\,C(z,\,\theta)\,G|0\rangle\otimes
B(z,\,\theta)\,G|0\rangle =0.
\]
\end{itemize}
One of the purposes of this paper is to establish the field theoretic
description of the SKP hierarchies. Instead of employing the bilinear
identity mentioned above, we will follow another route to the
goal. Namely we will translate the frame matrix description of the
{\sG} given in
\cite{ike-yama} into the language of the B-C system. At first this
route might seem somewhat roundabout but it has the merit of
being explicit and concrete.
When one tries to constract the super-analogue of the
\(\tau\)-function in the SKP hierarchies, he immediately encounters
a problem of (non-matching) time variables. It seems the time
variables which parameterize the flows do not match any
(super-)bosonization scheme of the B-C system. For example consider
the well known bosonization of the
\(\beta\)-\(\gamma\) system in Ref.\cite{fms}.
The current of the additional fermionic pair of the b-c type usually
denoted by \((\xi,\,\eta)\), which is necessary for the bosonization,
can not be expressed as a bilinear form of the fields \(\beta\) and
\(\gamma\).  Consequently the oscillators of this current can not be
identified with the time evolution operators on the {\sG}. Moreover
simple counting of the freedom indicates the number of the time
variables of the MRSKP or JSKP hierarchy is just half of the freedom
necessary for (super-)bosonizing the B-C system. These observation
indicates the necessity of a new SKP hierarchy. We recognize that if
one likes to formulate the SKP hierarchy as a dynamical system on
the {\sG}, the time evolution operators should be constructed from
currents in the bilinear form of the fields \(B\) and \(C\). The
possible combinations of such currents are given by
\(j^+=-bc\), \(j^-=-\beta\gamma\), \(\psi^+=c\beta\) and
\(\psi^-=\gamma b\), which form a set of generators of the Lie
superalgebra \(\gl11\). Actually one finds that the flows of the MRSKP
(JSKP) hierarchy are generated by the positive frequency part only of
the currents \(j^{tot}=-(bc+\beta\gamma)\) and \(\psi=c\beta-\gamma
b\) (\(j^{tot}\) and \(\psi^+\)). A superbosonization of the B-C
system (the super-Weyl algebra in their terminology) based on the the
Lie superalgebra \(\gl11\) has been found by Kac and van de
Leur\cite{kac-vdleur}. This enable us to consider a maximal SKP
hierarchy whose time flows are governed by all the above \(\gl11\)
currents, \(j^\pm\) and \(\psi^\pm\). In this way the MRSKP and JSKP
hierarchies are naturally extended. It should be mentioned that Rabin
also considered this type of a 'maximal' SKP hierarchy by connecting
MRSKP and JSKP to the superbosonization of Kac-van de Leur.  However,
{\em various ways appropriately parameterizing the extended flows
without spoiling the integrability} do not seem to be considered
seriously.  Our proposal for the time evolution operator compatible
with integrability is given by (\ref{opevol}). It seems to give the
simplest coordinatization of the extended flows. The resulting SKP
hierarchy is shown in (\ref{maxSKP}) in the form of the Sato
equations.  Then we can proceed as in the case of the KP
hierarchy. The \(\tau\)-function can be defined, from which the wave
function and its dual are derived acting the appropriate vertex
operators corresponding to the field
\(C\) and \(B\).  In the viewpoint of the superbosonization, we have
realized (the fixed total charge sector of) the B-C system on the
graded polynomial ring \(\complnum[t^\pm_1,t^\pm_2,\cdots]\) (where
\(t^\pm_{2n}\)'s are even and \(t^\pm_{2n+1}\)'s are odd variables),
although our vertex operators are somewhat complicated. In other words
{\em we have constructed the maximal SKP hierarchy which unifies the
known SKP hierarchies} at the cost of the simplicity of the
corresponding superbosonization. What extent of this construction of
the SKP hierarchy is considered to be natural is a remaining
question.\\
\indent
We have succeeded in the first goal of the program for the
Grassmannian approach to the SKP hierarchy.  There are many subjects
left for future investigations. One can think of, for example,
the problems listed at the beginning, regarding the maximal SKP hierarchy.
We have said virtually nothing about the geometrical aspects of the
theory. We expect, however, that the field theoretic description given
in this paper will shed light on our geometrical understanding of the
SKP hierarchies, combining our knowledge of the conformal field
theories on general (super-)Riemann surfaces\cite{alvarez-g}.  (We
will make some comments in the last section.)\\
\indent
The article organized as follows.  In {\S} 2, arranging the materials
necessary for the later use, we deal with the super-Sato
correspondence\cite{mula} between the space of the wave
(super-pseudodifferential) operators and a certain {\sG}, along the
arguments in Refs.\cite{mula} and \cite{ike-yama}.  In {\S} 3, we
describe of the known SKP hierarchies in the form of the
Sato equation. Then the interpretation of the hierarchies as dynamical
systems on the {\sG} becomes clear. With these preliminaries we
present the maximal SKP hierarchy.  The notion of the wave function
and its dual wave function are introduced and their characterization
through the bilinear identity is given in a general form.  In {\S} 4,
we introduce the B-C system. Translating the frame matrix
representation of the {\sG} into the language of the B-C system we
express the basic ingredients, i.e., the wave function, the dual wave
function and the \(\tau\)-function, in terms of them.  We see the free
fermion description of the KP hierarchy can be generalized in a very
natural way to the super case.  In {\S} 5, We briefly review the
superbosonization of the Kac and van de Leur and then, using this as an
intermediate step, construct the vertex operators corresponding to
the fields \(B\) and \(C\). In the last section, we mention the
geometrical side of the theory together with a remark on another SKP
hierarchy given by LeClair\cite{leclair}.

\resection{Super Sato correspondence}
To describe the {\skphs} as a dynamical system on the {\sG}
manifold, we first recall the super-Sato correspondence in some
detail.  Our exposition here is mostly due to \cite{mula} and
\cite{ike-yama}.  Let us start with the description of
the basic ingredients of the super-pseudo differential
operators({\spdos}). For more mathematical details see \cite{mula}.
Let \(\ala\) be some superalgebra modeled over \(\complnum\).  (We do
not specify its precise content. The simplest case is that \(\ala\)
is
\(\complnum\) itself.) We define as our function space the
supercommutative algebra
\beq
R=\ala \mathop{\otimes}_\complnum \ring.
\eeq
Here \(\ring \) is the ring of formal power series in even variable
\(x\) and odd nilpotent variable \(\xi\) that satisfy
\( x\xi=\xi x\) and \(\xi^2=0\). The \(\bZ_2\) gradation is
introduced naturally in \(R\) from those of \(\ala\) and \(\ring\) :
 \(R=R_0\oplus R_1\).
\(R\) has a super-derivation operator \(D=\der-{\xi} +\xi\der-x\) satisfying
\(D^2=\der-x\). We define the algebra \(\E\) of {\spdos} by
\beq
\E=R((\pd[j])) = \big\{\> P=\op[p,({j< \infty};j)],\quad p_j\in R\>\big\},
\eeq
where $\pd[1]=\pder[1]-x D$. Let \(\D\) be the subalgebra consisting
of super-differential operators:
\beq
\D=R[D]=\big\{\>P=\op[p,({0\leq j< \infty} ;j)] \>\big\}\subset \E.
\eeq
Then, we have the splitting of \(\E\):
\beq
\E=\D\oplus\E_{-},
\eeq
where
\beq
\E_{-}=\pd[1]R[[\pd[1] ]]=\big\{\;P=\op[p,({j \geq 1};{-j})]\;\big\}.
\eeq
The algebraic structure of \(\E\) is introduced through the
super-Leibniz rule. Let \(f\) be a homogeneous element (i.e. with
fixed Grassmann parity) of \(R\).
Then we have
\beq
D^m f=\sum_{r=0}\;{m\brack r}(-1)^{|f|(m-r)}f^{[r]}D^{m-r},
\eeq
where \( |f|\) denotes the Grassmann parity of \(f\),
\(f^{[r]}\) denotes the r-th derivative of \(f\) with respect to
\(D\) and \({m\brack r}\) are the super-binomial
coefficients\cite{man-rad} defined by
\beq
{m\brack r}=\left\{
\begin{array}{cl}
0,&\quad \mbox{for $r(m-r)=1\quad {\rm mod}\>\> 2$}\\
\left({{\gauss(m,2)}\atop {\gauss(r,2)}}\right),
&\quad \mbox{for $r(m-r)=0\quad
{\rm mod}\>\> 2$\quad .}
\end{array}\right.
\eeq
As a special subalgebra of \(\E\), we define \(\sgrass\) that consists of the
homogeneous even monic {\spdos}:
\beq
\sgrass = \big\{\; S\in \E\;\bigr|\; S=\sop,\quad
|s_j|= \left\{\scriptstyle{
\begin{array}{cl}
0,&\quad\mbox{for \(j\) even}\\
1,&\quad\mbox{for \(j\) odd}
\end{array}}\right.\>\big\}
\eeq
\(\sgrass\) has a group structure (i.e super-Volterra group) since
all elements of \(\sgrass\) are invertible in \(\sgrass\).

Next, following Mulase\cite{mula}, we give a definition of the
{\sG} and derive a supersymmetric generalization
of the theorem of Sato, which states the relation between the
{\spdos} and the {\sG}. Let us introduce
a new pair of even and odd variables \(z\) and \(\theta\)
which are regarded as Fourier transformations of
\(\der-x\) and \(\der-{\xi}\) respectively.
Furthermore it is useful to introduce symbols \(\zeta^m,\>\; m\in\bZ\)
\cite{mula} defined by
\beq
\left\{
\begin{array}{rl}
\zeta^{2m}&=z^m\\
\zeta^{2m+1}&=z^m\theta .
\end{array}\right.
\eeq
We define the super-linear space (i.e. free \(\ala\)-module) \(\H \)
as the space of the formal Laurent series :
\beq
\H=\complnum((\;z^{-1},\theta\;))\otimes\ala=
\big\{\;f=\sum_{n< \infty}\zeta^n f_n,\quad f_n\in\ala \;\big\}.
\eeq
As in the case of \(\E\), the superspace structure (i.e.
\(\bZ_2\)-gradation) of \(\H\) is obvious. There is a natural
direct sum decomposition of \(\H\):
\beq
\H=\H_+ \oplus \H_-,
\eeq
where \(\H_+\) and \(\H_-\) are the subspaces defined by respectively
\beq
\H_+=\complnum[z,\theta]\otimes\ala =
\big\{\;f=\sum_{0\leq n< \infty}\zeta^n f_n\;\big\},
\eeq
and
\beq
\H_-=z^{-1}\complnum[[z^{-1},\theta]]\otimes\ala =
\big\{\;f=\sum_{n\geq 1}\zeta^{-n} f_{-n}\;\big\}.
\eeq
Let \(\pi_+\) be the projection \(\pi:\,\H\longrightarrow \H_+\).
The {\sG} concerned in this paper, which is denoted by \(\G\),
is defined as a set of subspaces of \(\H\) such that
\beq
\G=\big\{\;W\subset \H |\;W\oplus \H_- =\H\;\big\}.
\eeq
In other words, the subspace \(W\subset \H\) belongs
to \(\G\) if and only if the projection of each subspace \(W\) to \(\H_+\)
is bijective, i.e. \({\rm ker}\;\pi_+\bigr|_W=0\) and
\({\rm coker}\;\pi_+\bigr|_W=0\). Of course this definition of the
{\sG} is a very restricted one corresponding to a supersymmetric
generalization of the Sato Grassmannian, and see Mulase\cite{mula} for
more general definition of the {\sG s} with an arbitrary level and
Fredholm index. \(\G\) is then viewed as a big cell of the {\sG}
with Fredholm index \(0|0\).\par
The super-linear space \(\H\) can be viewed as an \(\E\)-module as
follows. Having in mind the Fourier transform, we convert
\( x,\>\,\xi,\;\,\der-x\) and \(\der-\xi\) to
\(-\der-z,\;\,\der-\theta,\;\,z\) and \(\theta\), and consider
a homomorphism \(\rho\) from \(\E\) into the space of differential
operators on \(\H\):
\beq
\rho (P)=\sum_{j<\infty}\big\{\>
p_{2j}(-\der-z,\der-\theta)\,z^j + p_{2j+1}(-\der-z,\der-\theta)\,
z^j(\theta + z\der-\theta)\>\big\},\quad
P=\sum_{j<\infty}p_j(x,\xi)D^j\in\E. \la{rho}
\eeq
Through the homomorphism \(\rho\), {\spdos} act on \(\H\) and hence \(\H\)
is considered as an \(\E\)-module. Furthermore, we obtain a super-linear
transform from \(\E\) to \(\H\), which is denoted by \(\tilrho\), by setting
\beq
\tilrho(P)=\rho(P)\cdot 1\>\in \H, \quad P\in \E.
\eeq
\begin{th}
There exists a one to one correspondence between the {\sG} \(\G\)
and \(\sgrass\).
The bijection \(\sigma:\;\sgrass\> \widetilde{\longrightarrow}\> \G\)
is obtained by associating each \(S\) in \(\sgrass\)
with a subspace \(\tilrho (S^{-1}\D)\subset \H\) :
\beq
\sgrass\ni S\mathop{\longmapsto}^{\sigma}\;\tilrho(S^{-1}\D)\in\G\>.
\eeq
\end{th}
First we must show that the well-difinedness of the map \(\sigma\), i.e.
\(\sigma(S)\in \G\). From the expression (\ref{rho}) of the
homomorphism \(\rho\) we note that
\begin{enumerate}
\item
\(\tilrho(\D)=\H_+\)
\item
\(\rho(S)\H=\H,\quad S\in\sgrass\)
\item
\(\rho(S)\H_-=\H_-, \quad S\in \sgrass\).
\end{enumerate}
Then we have
\(\H=\rho(S^{-1})\H=\rho(S^{-1})\H_+\oplus \rho(S^{-1})\H_-=
\tilrho(S^{-1}\D)\oplus\H_-\) for \(S\in\sgrass\), and
hence \(\sigma(S)=\tilrho(S^{-1}\D)\in\G\). To see that \(\sigma\)
is a bijection, we need to investigate the homomorphism \(\rho\) in more
detail. Let \(P\) be a homogeneous element of \(\E\) and express it as
\beqn
P&=&\sum_{j<\infty}p_j(x,\xi)D^j,\nonumber\\
p_j(x,\xi)&=&\sum_{n=0}\frac{x^n}{n!} (a_{2n,j}+\xi a_{2n+1,j}),\quad
|a_{m,j}|=m+j+|P|,\>\> {\rm mod}\>\;2.
\eeqn
Then we set
\beq
v_P^{(m)}=\rho(P)\zeta^m=\sum_j \zeta^j F[P]_{jm},\quad m\in \bZ
\la{base}
\eeq
and consider the coefficients \(F[P]_{jm}\) as matrix elements
of a \(\bZ\times \bZ \) matrix \(F[P]\).
By calculating
\beqn
v_P^{(m)}&=&\rho(P)\zeta^m=\tilrho(PD^m)\nonumber \\
&=&\sum_{j<\infty}\sum_{n=0}\>\frac{1}{n!}(-\der-z)^n
(a_{2n,j}+\der-\theta\, a_{2n+1,j})\,\zeta^{j+m},
\eeqn
we obtain the following result:
\beq
F[P]_{jm}=(-1)^{j+m+|P|}\,\Xi[P]_{-(j+1),-(m+1)},\quad j,m\in \bZ,
\la{frame}\eeq
with
\beq
{\widetilde \Xi}[P]_{jm}=
\sum_{k=0}{j \brack k}\;{\tilde a}_{k,m-j+k},
\qquad
\hbox{where}\quad
\left\{
\begin{array}{lcl}
{\tilde a}_{j,k}&=&(-1)^{(j+|P|)(k+1)}\,a_{j,k}\\
&&\\
{\widetilde \Xi}[P]_{jk}&=&(-1)^{(j+|P|)(k+1)}\,\Xi[P]_{jk}.
\end{array}\right.
\la{framexi}
\eeq
We write (\ref{frame}) in the matrix form
\beq
F[P]=(-1)^{|P|}\,JK\Xi[P] KJ,\la{framat}
\eeq
where \(J\) and \(K\) are the matrices defined by
\beq
J=(\;(-1)^j\,\delta_{jk}\;)_{j,k\in\bZ}\quad\hbox{and}\quad
K=(\;\delta_{j+k+1,0}\;)_{j,k\in\bZ}.
\eeq
Using the properties of the super-binomial coefficients, the
expression (\ref{framexi}) can be inverted as
\beq
{\tilde a}_{j,m}=\sum_{k=0}^j{j \brack k}
(-1)^{\frac{j(j+1)-k(k+1)}{2}}\;{\widetilde \Xi}[P]_{k,m-j+k},\qquad
j\in\bZzp,\quad m\in\bZ. \la{coff-a}
\eeq
We see that for an arbitrary integer \(p\),
\( (a_{j,m})_{j\in\bZzp,\>\> m\in\> p+\bZm}\) and
\( (\Xi[P]_{jm})_{j\in\bZzp,\>\> m\in\> p+\bZm}\) can be expressed each
other by the above transforms.
\begin{lm}\la{lem1}
If a {\spdo} \(P\) preserves \(\H_+\), i.e.
\(\rho (P)\H_+ \subset \H_+,\>\; P\in\E\), then \(P\) must be
a differential operator, i.e. \(P\in\D\).
\end{lm}
[{\it Proof}]\quad The condition \(\rho (P)\H_+\subset \H_+\) means
\(v_P^{(m)}\in\H_+,\>\>m=0,1,2\cdots\)\quad and so
\(
F[P]_{jm}=0,\quad \hbox{for }j\in \bZm,\>\;m\in\bZzp
\),
( equivalently
\(\Xi[P]_{jm}=0\), for \(j\in \bZzp,\>\;m\in\bZm\)).
Then it follows from (\ref{coff-a}) that \(a_{jm}=0,\hbox{ for }
j\in\bZzp,\>\;m\in\bZm\), showing \(P\in\D\).\hfill$\Box$\\
\indent Let \(P\in \sgrass\) then in (\ref{rho})
\beq
\begin{array}{rl}
a_{jm}=0,&\quad \hbox{for }m>0,\\
a_{j0}=0,&\quad \hbox{for }j>0,\hbox{ and }a_{00}=1.
\end{array}
\eeq
In terms of \(\Xi[P]\) we find
\begin{enumerate}
\item
\(\Xi[P]_{jm}=\delta_{jm}\) for \(j\leq m\)\quad(i.e. \(\Xi[P]\) becomes
a triangular matrix with unit diagonal elements.)
\item
In specifying \(P\in\sgrass\) we are free to set the values
\((\Xi[P]_{jm})_{j\in\bZzp,\>m\in\bZm}\) ( equivalently
\((a_{j,m})_{j\in\bZzp,\>m\in\bZm}\)).
The other elements of the triangular matrix \(\Xi[P]\) are determined
by that part of the elements as
\beqn
\Xi_{jm}&=&\sum_{k=0}^{j-m}f(j,m)_k\,\Xi_{k,k-(j-m)}\qquad\hbox{for}
\quad j\geq m>0 \la{constr1}\\
\Xi_{-(j+1),-(m+1)}&=&\sum_{k=0}^{m-j}h(j,m)_k\,\Xi_{k,k-(m-j)}
\qquad\hbox{for}\quad m\geq j\geq 0\la{constr2}
\eeqn
We do not need to know the explicit forms of the coefficients
\(f(j,m)_k\) and \(h(j,m)_k\) in the following.
\end{enumerate}
Conversely, we have the following lemma.
\begin{lm}
\la{lem2}
A triangular matrix \((\Xi_{ij})\) with \(\Xi_{ij}=\delta_{ij}\) for
\(j\leq m\) is associated with some {\spdo} \(S\in\sgrass\)
if and only if it satisfies the relations (\ref{constr1}) and
(\ref{constr2}) among its elements.
\end{lm}
[{\it Proof}]\quad By (\ref{coff-a}),
\((\Xi_{jm})_{j\in\bZzp,\>m\in\bZm}\) determines
\((a_{j,m})_{j\in\bZzp,\>m\in\bZm}\) and so \(S\in\sgrass\).
\hfill$\Box$\\
\vskip 2mm \indent
Now we return to the theorem.
\begin{enumerate}
\renewcommand{\labelenumi}{(\roman{enumi})}
\item Injectivity of \(\sigma\)\\
Suppose that \(\sigma (S_1)=\sigma (S_2)\). Then we have
\[
\tilrho(S_1S_2^{-1}\D)=\tilrho(\D)=\H_+.
\]
{}From lemma~\ref{lem1}, it follows
\(S_1S_2^{-1}\in\D\cap\sgrass=\{1\}\), thus we have \(S_1=S_2\).
\item Surjectivity of \(\sigma\)\\
Let \(W\)  be an arbitrary element of \(\G\). To prove the
surjectivity, it is sufficient to find a basis for \(W\) of the form
\beq
v^{(m)}(\zeta)=\sum_{k\leq m}(-1)^{k+m}\zeta^k\Xi_{-(k+1),-(m+1)},
\qquad m=0,1,2,\cdots\la{basis}
\eeq
where the set of coefficients \((\Xi_{jk})_{j\geq k,\>k\in\bZm}\)
can be considered as a part of a triangular matrix that
satisfies the relations (\ref{constr1}) and (\ref{constr2}).
Then Lemma~\ref{lem2} indicates \(\Xi  =\Xi[S]\) with some
\(S\in\sgrass\) and (\ref{basis}) is nothing but
\(\tilrho(S\D)=\sigma (S^{-1})=W\), which shows the surjectivity
of \(\sigma\). Let \(\{w^{(m)}\},\>\;m=0,1,2,\cdots\) be the canonical
basis for \(w\) which takes the form
\beq
w^{(m)}(\zeta)=\zeta^m+\sum_{k>0}\zeta^{-k}w_{-k,m}.
\eeq
We can determine inductively the matrix elements \(\Xi_{jk}\)
from the coefficients \((w_{j,k})\) to satisfy the relation
(\ref{constr1}) and (\ref{constr2}).
First we set \(v^{(0)}=w^{(0)}\) which fixes \(-1\)st column
of \(\Xi\) as
\[
\Xi_{-1,-1}=1,\quad\hbox{and}\quad
\Xi_{k,-1}=(-1)^{k+1}w_{-(k+1),0}\quad k=0,1,2,\cdots .
\]
Suppose that we have determined the matrix \(\Xi\) up to \(-n\)-th column,
i.e. \(\Xi_{jk}\) with \(0>k\geq -n, \>\,j\geq k\).
Then \(\Xi_{-(j+1),-(n+1)},\>\;0\leq j\leq n\) can be fixed from
(\ref{constr2}), and we set
\[
v^{(n)}(\zeta)=w^{(n)}(\zeta)+(-1)^n
\sum_{j=1}\Xi_{-(n-j+1),-(n+1)}w^{(n-j)}(\zeta),
\]
which determine the elements \(\Xi_{j,-(n+1)}\>\;j\geq 0\).
The relation (\ref{constr1}) only states positive columns of \(\Xi\)
should be fixed by the negative ones. Now we complete the proof of
the theorem.
\end{enumerate}
Let \(\Xi\in \matZZA\).
We have observed in (\ref{framat}) that for a given \(S\in\sgrass\),
\(\Xi[S^{-1}]_{\ltinf}\) \footnote{
Throughout this paper,
the subscripts \(\leq\) (\(\geq\)) and \(<\) (\(>\)) indicate
taking the part of non-negative (non-positive) integers and
the part of negative (positive) integers, respectively.
For example, let \(\Xi\in \matZZA\), then \(\Xi\) is written in a block form
\[
\Xi=\left(\begin{array}{cc}
\Xi_{\wgeinf}&\Xi_{\geltinf}\\
\Xi_{\ltgeinf}&\Xi_{\wltinf}
\end{array}
\right).
\]
We further set
\beq
\Xi_{\geinf}=\left(
\begin{array}{c}
\Xi_{\wgeinf}\\
\Xi_{\ltgeinf}
\end{array}\right)\quad\hbox{and}\quad
\Xi_{\ltinf}=\left(
\begin{array}{c}
\Xi_{\geltinf}\\
\Xi_{\wltinf}
\end{array}\right).
\eeq}
defined through (\ref{framexi}) gives a frame
matrix for an associated subspace \(W=\sigma(S)\in\G\). There exists
a natural way to obtain \(\Xi[P]\) from a given {\spdo} \(P\in\E\).
Let us define a \(\bZ\times\bZ\) matrix \(\psi[P]=(\psi[P]_{jk})\),
\(P\in\E\) by
\beq
D^jP=\sum_k\psi[P]_{jk}D^k.
\eeq
then the  mapping \(\psi\)
\beq
\E\ni P\mathop{\longrightarrow}^\psi \psi[P]
\in \matZZR.
\eeq
becomes an injective algebra homomorphism and thus gives a matrix
representation of the algebra \(\E\). The following proposition
shows how the matrix \(\Xi[P]\) arises from \(\psi[P]\).
\begin{pro}
\la{pro1}
For a given {\spdo} \(P\), the matrix \(\psi[P]\) takes the form
\beq
\psi[P](x,\xi)=e^{\xximat}\;\Xi[P]\;e^{-(\xximat)},
\la{matpsdo}
\eeq
and hence \(\Xi[P]=\psi[P]\bigr|_{x=\xi=0}\), where \(\Xi[P]\)
is the matrix defined previously, and \(\Gamma\) and \(\Lambda\)
are the matrices given by
\beq
\Gamma=(\;\delta_{j+1,k}\;)_{j,k\in\bZ}\quad\hbox{\rm and}\quad
\Lambda=(\;\delta_{j+2,k}\;)_{j,k\in\bZ}=\Gamma^2.
\eeq
\end{pro}
It is useful to introduce a pairing
\(<\>,\> >:\H\times \H\rightarrow \ala\) defined by
\beq
<f,\,g>=\oovpi \oint_0 dzd\theta f(\zeta)g(\zeta),\qquad
f,\,g\in\H \la{pairing}
\eeq
\(<\>,\> >\) has the properties
\begin{eqnarray*}
<\lambda f,\,g>&=&(-1)^{|\lambda|}<f,\,g>,\qquad
<f,\,g\lambda>=<f,\,g>\lambda,\\
<f\lambda,\,g>&=&<f,\,\lambda g>,\qquad\lambda\in\ala,\\
<\zeta^m,\,\zeta^n>&=&\delta_{n+m+1,0}.
\end{eqnarray*}
To prove the proposition, let us consider the following integral
\beq
\oovpi\oint dzd\theta D^ke^{\xxi}\rho(P)\zeta^m
=<D^ke^{\xxi},\,\rho(P)\zeta^m>,\quad P\in\E.
\la{intf}
\eeq
Here we note the following identities:
\beqn
<D^ke^{\xxi},\,\zeta^j>&=&<e^{\xxi},\,\zeta^{k+j}>\nonumber \\
&=&\left\{
\begin{array}{cll}
\frac{x^n}{n!}&\quad\hbox{for}\>\;k+j+1=-2n,&n\in\bZzp\\
\frac{-\xi x^n}{n!}&\quad\hbox{for}\>\;k+j+1=-(2n+1),&n\in\bZzp\\
0&\quad\hbox{otherwise}\\
\end{array}\right. \nonumber \\
&=&(e^{x\Lambda-\xi\Gamma})_{k,-(j+1)}
=(e^{x\Lambda-\xi\Gamma}K)_{k,j},
\eeqn
from which it follows that
\beq
D^ke^{\xxi}=\sum_j(e^{\xximat})_{kj}\,\zeta^j\>.\la{Dexp}
\eeq
In addition we see for \(P\in\E\), \(f\in\H\), integrating by part
\beq
<e^{\xxi},\,\rho (P)f>=<Pe^{\xxi},\,f>.
\eeq
Making use of these equalities, we can calculate the expression
(\ref{intf}) in two ways. On the one hand, using the notation
(\ref{base}), we have
\begin{eqnarray*}
<D^ke^{\xxi},\,\rho (P)\zeta^m>&=&\sum_j
<D^ke^{\xxi},\,\zeta^j>F[P]_{jm}\\
&=&(e^{x\Lambda-\xi\Gamma}KF[P])_{km}.
\end{eqnarray*}
On the other hand, from the definition of \(\psi[P]\),
we have
\begin{eqnarray*}
<D^ke^{\xxi},\,\rho (P)\zeta^m>&=&<D^kPe^{\xxi},\,\zeta^m>\\
&=&\sum_j(-1)^{k+j+|P|}\psi[P]_{kj}<D^je^{\xxi},\,\zeta^m>\\
&=&(-1)^{|P|}(J\psi[P]Je^{x\Lambda-\xi\Gamma}K)_{km}.
\end{eqnarray*}
As a result we obtain
\beq
\psi[P]Je^{x\Lambda-\xi\Gamma}K=
(-1)^{|P|}Je^{x\Lambda-\xi\Gamma}KF[P].
\eeq
Putting (\ref{framat}) into the above formula and taking into account
of the matrix identities \(J\Gamma+\Gamma J={\bf 0}=JK+KJ\) and
\(K^2=J^2={\bf 1}\), we get (\ref{matpsdo}).\\ \indent
Lastly, we introduce the ``adjoint'' operation
\(\ast\) in \(\E\),
which plays an important role in the field theoretic
representation of the SKP hierarchy, and consider the
corresponding formula in the matrix representation \(\psi\).
The adjoint operation is an involution which is defined
uniquely from
\begin{enumerate}
\renewcommand{\labelenumi}{(\roman{enumi})}
\item \(D^*=-D\),
\item \(f^*=f\),\quad for \(f\in R\),
\item \((P_1P_2)^*=(-1)^{|P_1||P_2|}P_2^*P_1^*\)\quad for
any two homogeneous elements \(P_1\), \(P_2\in\E\).
\end{enumerate}
\begin{pro}
\la{pro2}
Let \(P\in\E\), then
\beq
\psi[P^*]=(-1)^{|P|}IK\psi[P]^{\rm st}KI,
\la{matadj}
\eeq
where \(I\) is a diagonal matrix given by
\beq
I=(\;(-1)^{\frac{j(j+1)}{2}}\delta_{jk}\;)_{j,k\in\bZ},
\eeq
and \(\psi[P]^{\rm st}\) is the supertransposition of \(\psi[P]\).
\end{pro}
The definition of the supertransposition (we adopted in this paper)
is as follows. Let \(X=(X_{jk})_{j,k\in\bZ}\) be a homogeneous
element of \(\matZZR\), i.e. \(|X_{jk}|=|X|+j+k,\>\;\hbox{mod}2\).
Then we set
\beq
(X^{\rm st})_{jk}=(-1)^{(|X|+k)(j+k)}X_{kj}.
\la{stransp}
\eeq
The supertransposition (\ref{stransp}) is defined so as to
satisfy
\beq
(XY)^{\rm st}=(-1)^{|X||Y|}Y^{\rm st}X^{\rm st}
\eeq
for all homogeneous \(X,\>Y\in\matZZR\). Note that successive
operations give
\beq
(X^{\rm st})^{\rm st}=JXJ.
\eeq
Now Proposition~\ref{pro2} follows from the definition of \(\psi\).
Taking the adjoint of
\[
D^jP=\sum_k\psi[P]_{jk}D^k,\qquad P\in\E,
\]
we get
\[
(-1)^{\frac{j(j+1)}{2}+|P|j}P^*D^j=
\sum_k (-1)^{\frac{k(k+1)}{2}+(|P|+k+j)k}D^k\psi[P]_{jk}.
\]
Applying \(D^m\), we further obtain
\begin{eqnarray*}
\lefteqn{(-1)^{\frac{j(j+1)}{2}+|P|j}\sum_l\psi[P^*]_{ml}D^{l+j}
=\sum_k(-1)^{\frac{k(k+1)}{2}+(|P|+k+j)k}D^{m+k}\psi[P]_{jk}}\\
&=&\sum_k\sum_{r=0}^{m+k}(-1)^{\frac{k(k+1)}{2}+(|P|+k+j)(m-r)}
{m+k \brack r}\psi[P]_{jk}^{[r]}D^{m+k-r}.
\end{eqnarray*}
Comparing the coefficients on both sides, we see
\begin{eqnarray*}
\lefteqn{
(-1)^{\frac{j(j+1)}{2}+|P|j}\psi[P^*]_{ml}}\\
&=&
\sum_{r=0}(-1)^{\frac{(l+j+r-m)(l+j+r-m+1)}{2}+(|P|+l+r-m)(m-r)}
{l+j+r\brack r}\psi[P]_{j,l+j+r-m}^{[r]}.
\end{eqnarray*}
Putting \(j=-(l+1)\) and taking account of
\({r-1\brack r}=\delta_{r,0}\), we get
\beq
\psi[P^*]_{ml}=(-1)^{\frac{(l-m)(l-m+1)}{2}+|P|(m+l+1)}
\psi[P]_{-(l+1),-(m+1)},
\eeq
which gives the matrix identity (\ref{matadj}).\\\indent
The following facts will be useful later on.
\begin{pro}
\la{pro3}
Let \(P,\>Q\in\E\), then\\
{\it (i)}
\beq
P^\ast e^{-(\xxi)}=\rho(P)\, e^{-(\xxi)}.
\eeq
{\it (ii)} \(P\in\D\) if and only if
\[
<e^{\xxi},\,\rho (P)\zeta^m>=0,\qquad m=0,1,2,\cdots,
\]
or equivalently
\beq
<e^{\xxi},\,\rho (P)e^{-(x'z+\xi'\theta)}>=0.
\eeq
{\it (iii)} \(PQ^\ast\in\D\) if and only if
\beq
<P(x,\xi)e^{\xxi},\,Q(x',\xi')e^{-(x'z+\xi'\theta)}>=0.\la{pro32}
\eeq
\end{pro}
[{\it Proof}]\quad
Let \(P=\sum_{j<\infty}p_j(x,\xi)D^j\).
(i) follows from a straightforward computation.
\beqn
P^\ast e^{-(\xxi)}&=&\sum_j(-1)^{\frac{j(j+1)}{2}+(j+|P|)j}
D^jp_j(x,\xi)\,e^{-(\xxi)}
\nonumber\\
&=&\sum_j(-1)^{\frac{j(j+1)}{2}+(j+|P|)j}
D^jp_j(-\der-z,\der-\theta)\,e^{-(\xxi)}
\nonumber\\
&=&\sum_j(-1)^{\frac{j(j+1)}{2}}p_j(-\der-z,\der-\theta)D^j\,e^{-(\xxi)}
\nonumber\\
&=&\sum_j p_j(-\der-z,\der-\theta)(\theta+z\der-\theta)^j\,e^{-(\xxi)}
\nonumber\\
&=&\rho(P)\, e^{-(\xxi)}.
\eeqn
(ii) is simply a paraphrase of Lemma~\ref{lem1} since
for \(f(\zeta)\in\H\), \(<e^{\xxi},\>f(\zeta)>=0\) means
\(f(\zeta)\in\H_+\) and hence (\ref{pro32}) is a statement
\(\rho (P)\H_+\subset \H_+\).\hfill\break
(iii) follows from (i) and (ii).\hfill$\Box$

\resection{Time evolution}
\subsection{Time evolution and the known SKP hierarchies}
The super-Sato correspondence leads us to the natural interpretation
of SKP hierarchies as dynamical system on the {\sG} \(\G\).
As stated in Introduction, various SKP hierarchies emerge
according to various ways of introducing the set of infinitely
many (even and odd) flows on the \(\G\).
First we start to describe the two known SKP hierarchies, i.e.
MRSKP and JSKP hierarchies. Let \(\{t_n\}_{n\geq 1}\) be the set of
an infinite number of the time variables, where the even times
\(\{t_{2n}\}_{n\geq 1}\) are even variables and odd times
\(\{t_{2n-1}\}_{n\geq 1}\) are odd variables, respectively.
Now we define the time evolution operators \(\evol[MR]\)
and \(\evol[J]\)
\beq
\begin{array}{rlll}
\evol[MR]&=e^{\hamil[MR]}, \qquad&\hbox{with}&\quad
\hamil[MR]=\sum_{n=1}t_nD^n,\\
&&&\\
\evol[J]&=e^{\hamil[J]}, \qquad&\hbox{with}&\quad
\hamil[J]=\sum_{n=0}\big(t_{2(n+1)}{\der-x}^{n+1}
+t_{\podd-n}{\der-x}^n\der-\xi\big).\la{3.1}
\end{array}
\eeq
\(\evol[MR]\) and \(\evol[J]\) act on \(\G\) through
\(\rho(\evol[MR,\>J])\G\) and define the flows for MRSKP and JSKP
hierarchies. In introducing the infinitely many time variables,
we should extend the super algebra \(\ala\) to
\(\ala[[t_1,t_2,\cdots]]\) and also extend the definition
of \(\E\), \(\E_-\), \(\D\) and \(\sgrass\) appropriately to
accommodate the time dependence.  (See, for example, the prescription
in reference
\cite{mula}.)  Here, we abuse
the same notation \(\E\), \(\E_-\), \(\D\) and \(\sgrass\) allowing
them to have the time dependence.  Assume \(U(t)=e^{H(t)}\) (with
\(H=\hamil[MR]\) or \(\hamil[J]\)) generates the flows on \(\G\). It
means that, for an arbitrary element
\(W=\rho(S_0^{-1})\H_+\) we have
\(\rho\big(U(t)\big)W=\rho\big(U(t)S_0^{-1}\big)\H_+\in\G\).
Then, from the theorem in the previous section, there exists
a {\spdo} \(S(t)\in\sgrass\) such that
\beq
\rho\big(U(t)S_0^{-1}\big)\H_+=\rho\big(S^{-1}(t)\big)\H_+.
\eeq
Thus \(S(t)U(t)S_0^{-1}\) preserves \(H_+\), and hence from
Lemma~\ref{lem1} we have
\beq
\big(S(t)U(t)S_0^{-1}\big)_-=0.\la{grasseq}
\eeq
We refer Eq.(\ref{grasseq}) as the Grassmann equation according to
\cite{ike-yama}. In \cite{mula2} Mulase established the supersymmetric
generalization of the Birkhoff decomposition, which states
the unique solvability of Eq.(\ref{grasseq}), in other words,
\(U(t)S_0^{-1}\) is factorized uniquely into the form
\beq
U(t)S_0^{-1}=S(t)^{-1}Y(t), \qquad S(t)\in \sgrass,\quad
Y(t)\in \D \quad\mbox{s.t.}\quad Y(0)=1.\la{bkffact}
\eeq
Note that the differential operator \(Y(t)\) is invertible in \(\D\).
The operator \(S(t)\) satisfies the system of the equations
(i.e. the SKP hierarchy) derived from (\ref{grasseq}).
Let us define a one form
\beq
\Omega\equiv dU(t)U(t)^{-1}=\sum_{n=1}dt_n\Omega_n,\la{omega}
\eeq
where \(d=\sum_{n=1}dt_n\der-{t_n}\). The explicit forms
of the operators \(\Omega_n\) are
\beq
\begin{array}{ll}
{\left\{\begin{array}{ll}
\Omega_{2n}&={\der-x}^n,\\
\Omega_{\podd-n}&={\der-x}^n\der-{\xi}.
\end{array}\right.}&\qquad
\hbox{for JSKP}\\&\\
{\left\{\begin{array}{ll}
\Omega_{2n}&={\der-x}^n,\\
\Omega_{\podd-n}&=D^{\podd-n}(1-\sum_{m=0}t_{\podd-m}D^{\podd-m}),
\end{array}\right.}&\qquad
\hbox{for MRSKP}
\end{array}
\eeq
Applying \(d\) to (\ref{bkffact}) and decomposing it into the \(\E_-\)
part and \(\D\) part, we have
\beqn
dS&=&-(S\Omega S^{-1})_-\,S,\la{skp}\\
dY&=&(S\Omega S^{-1})_+\,Y.\la{skpplus}
\eeqn
Equation (\ref{skp}) is equivalent to the system
\beq
\left\{\begin{array}{rcl}
\der-{t_{2n}}S&=&-(S{\der-x}^nS^{-1})_-S,\\
\der-{t_{\podd-n}}S&=&-(S{\der-x}^n\der-\xi S^{-1})_-S,
\end{array}\right.
\qquad \mbox{for JSKP}
\la{satoeqJ}
\eeq
and
\beq
D_nS=-(SD^nS^{-1})_-S,\qquad\mbox{for MRSKP}
\la{satoeqMR}
\eeq
where
\beq
\left\{\begin{array}{rcl}
D_{2n}&=&\der-{t_{2n}},\\
D_{\podd-n}&=&\der-{t_{\podd-n}}-\sum_{m=0}t_{\podd-m}\der-{t_{2(n+m+1)}}.
\end{array}\right.
\eeq
Here we mention the unique solvability of the initial value problem
for the system (\ref{satoeqJ}) and (\ref{satoeqMR}) proved by
Mulase in \cite{mula2}. The argument is outlined as follows.
To obtain a unique solution of (\ref{skp}), first solve
\beq
dZ=\Omega Z,\la{linprbm}
\eeq
where \(Z\) is a {\spdo} such that \(Z(0)\in\sgrass\).
The generalized Birkhoff decomposition theorem previously mentioned
gives unique factorization
\(Z(t)=S(t)^{-1}Y(t), \quad S(t)\in \sgrass\),
\(Y(t)\in \D\) s.t. \(Y(0)=1\).
Then \(S(t)\) and \(Y(t)\) solve (\ref{skp}) and (\ref{skpplus})
respectively, as viewed in the previous paragraph.
The uniqueness of the solution \(S(t)\) results from the
unique solvability (of the initial value problem) of the equation
(\ref{linprbm}) (at least when we restrict ourselves
\(\ala=\complnum\)) and the uniqueness of the factorization
of \(Z\).

\subsection{Maximal SKP hierarchy}
Looking closely at the form \(\evol[J]\) and \(\evol[MR]\), we now
present a new time evolution operator which generates all the flows of
JSKP and MRSKP hierarchies. This corresponds to the one introduced as
`maximal' SKP hierarchy in \cite{rabin}, so we will call the resultant
system by this name.  In order to realise the 'maximal' SKP hierarchy,
we have to double the number of time variables, i.e. we introduce two
infinite sets of the time variables
\(\{\pt-n\}_{n\geq 1}\) and \(\{\mt-n\}_{n\geq 1}\), where
even times \(\pmt-{2n}\) and odd times \(\pmt-{\podd-n}\) are
even and odd respectively as before.
The time evolution operator we consider is defined as
\beq
\evol[M]=\evol[(1)]\evol[(0)],
\eeq
where
\beqn
\evol[(1)]=e^{\hamil[(1)]}\quad&\hbox{with}&\quad
\hamil[(1)]=\sum_{n=0}
(\,\pt-{\podd-n}\der-\xi{\der-x}^n+\mt-{\podd-n}\xi{\der-x}^{n+1}\,),
\la{hamil1}\\
\evol[(0)]=e^{\hamil[(0)]}\quad&\hbox{with}&\quad
\hamil[(0)]=\sum_{n=0}
(\,\pt-{2(n+1)}\der-\xi\xi\,{\der-x}^{n+1}
+\mt-{2(n+1)}\xi\der-\xi\,{\der-x}^{n+1}\,).\la{hamil0}
\eeqn
Since \(\hamil[(1)]\) and \(\hamil[(0)]\) do not commute each other,
exponentiation of all the vector fields
\(\hamil[(1)]\) and \(\hamil[(0)]\) in a form such as
\(e^{\hamil[(1)]+\hamil[(0)]}\) turns out unsuccessful.
Instead we consider successive time evolutions: \(\evol[(0)]\) first
generates even flows and subsequently \(\evol[(1)]\) generates odd
flows.  From the definition, it is evident that the JSKP and the MRSKP
hierarchies are special cases of this maximal hierarchy and are
obtained by setting the time variables respectively,
\beqn
\pt-{2n}&=&\mt-{2n}=t_{2n},\quad \mt-{\podd-n}=0,\quad
\pt-{\podd-n}\rightarrow t_{\podd-n}\quad\hbox{for JSKP},\nonumber\\
\pt-{2n}&=&\mt-{2n}=t_{2n},\quad
\pt-{\podd-n}=\mt-{\podd-n}=t_{\podd-n}\quad\hbox{for MRSKP}.
\la{redtime}
\eeqn
As before we define the one form
\beq
\Omega_M=\sum_{n=1}(\,d\pt-n\Omega_n^+ + d\mt-n\Omega_n^-\,)
\eeq
by
\beq
\Omega_M\equiv d\evol[M]{\evol[M]}^{-1}
=d\evol[(1)]{\evol[(1)]}^{-1}+
\evol[(1)](\,d\evol[(0)]{\evol[(0)]}^{-1}\,){\evol[(1)]}^{-1},
\eeq
where \(d=\sum_{n=1}(\,d\pt-n\der-{\pt-n}+d\mt-n\der-{\mt-n}\,)\).
Then we have
\beqn
\Omega^+_\podd-n&=&\der-\xi{\der-x}^n
+\frac12 \sum_{m=0}\mt-{\podd-m}{\der-x}^{n+m+1},\nonumber\\
\Omega^-_\podd-n&=&\xi{\der-x}^{n+1}
+\frac12 \sum_{m=0}\pt-{\podd-m}{\der-x}^{n+m+1},\la{oddP}\\
\Omega^+_{2n}&=&\der-\xi\xi\,{\der-x}^n
-\sum_{m=0}(\,\pt-{\podd-m}\Omega^+_{\podd-{(n+m)}}-
\mt-{\podd-m}\Omega^-_{\podd-{(n+m)}}\,),\nonumber\\
\Omega^-_{2n}&=&\xi\der-\xi\,{\der-x}^n
+\sum_{m=0}(\,\pt-{\podd-m}\Omega^+_{\podd-{(n+m)}}-
\mt-{\podd-m}\Omega^-_{\podd-{(n+m)}}\,).\la{evenP}
\eeqn
The calculation of \(\Omega_M\) is carried out as follows.
First we note \({\hamil[(1)]}^3=0\), which follows from the
anti-commutativity of the odd times among themselves. We have thus
\beq
\evol[(1)]=1+\hamil[(1)]+\frac12 {\hamil[(1)]}^2,\quad
\hbox{and}\quad
{\evol[(1)]}^{-1}=1-\hamil[(1)]+\frac12 {\hamil[(1)]}^2,\quad
\eeq
Similarly, noting
\(\hamil[(1)]d\hamil[(1)]\,\hamil[(1)]=0\), we have
\beqn
\Omega_{(1)}&\equiv& d\evol[(1)]{\evol[(1)]}^{-1}\nonumber\\
&=&\left\{
d\hamil[(1)]+\frac12\left(d\hamil[(1)]\hamil[(1)]
+\hamil[(1)]d\hamil[(1)]\right)\right\}
\left(1-\hamil[(1)]+\frac12{\hamil[(1)]}^2\right)\nonumber\\
&=&d\hamil[(1)]+\frac12 [\hamil[(1)],\>d\hamil[(1)]].\la{omega1}
\eeqn
Putting into (\ref{omega1}) the expression (\ref{hamil1}) of
\(\hamil[(1)]\), we obtain (\ref{oddP}).
Next let us consider the even time part of \(\Omega\).
It is convenient to
separate \(\hamil[(0)]\) and \(\hamil[(1)]\) respectively
into two terms so that the each term contains only one type of time
variables (i.e. \(\pt-n\) or \(\mt-n\)):
\beqn
\hamil[(1)]&=&\hamil[(1)]^+ + \hamil[(1)]^-\quad\hbox{with}\quad
{\left\{\begin{array}{ll}
\hamil[(1)]^+&=\sum_{n=0}\pt-{\podd-n}\der-\xi{\der-x}^n,\\
\hamil[(1)]^-&=\sum_{n=0}\mt-{\podd-n}\xi{\der-x}^{n+1},
\end{array}\right.}\\
\hamil[(0)]&=&\hamil[(0)]^+ + \hamil[(0)]^-\quad\hbox{with}\quad
{\left\{\begin{array}{ll}
\hamil[(0)]^+&=\sum_{n=1}\pt-{2n}\;\der-\xi\xi\,{\der-x}^n,\\
\hamil[(0)]^-&=\sum_{n=1}\mt-{2n}\;\xi\der-\xi {\der-x}^n.
\end{array}\right.}
\eeqn
Because of
\(\hamil[(0)]^+\hamil[(0)]^-=\hamil[(0)]^-\hamil[(0)]^+=0\),
we have
\beq
\evol[(0)]=e^{\hamil[(0)]^+ + \hamil[(0)]^-}
=e^{\hamil[(0)]^+}+e^{\hamil[(0)]^-}-1
\eeq
and
\beqn
d\evol[(0)]{\evol[(0)]}^{-1}&=&
\left(d\hamil[(0)]^+e^{\hamil[(0)]^+}+
d\hamil[(0)]^-e^{\hamil[(0)]^-}\right)
\left(e^{-\hamil[(0)]^+}+e^{-\hamil[(0)]^-}-1\right)\nonumber\\
&=&d\hamil[(0)]
\eeqn
Furthermore let us write \(\hamil[(0)]\) as
\beq
\hamil[(0)]=E+{\bar E}\quad\hbox{with}\quad
\left\{\begin{array}{rl}
E&=\sum_{n=1}t_{2n}{\der-x}^n,\\
{\bar E}&=\sum_{n=1}{\bar t_{2n}}
(\der-\xi\xi-\xi\der-\xi){\der-x}^n,
\end{array}\right.
\eeq
where
\(t_{2n}\) and \({\bar t_{2n}}\) are defined as
\beq
t_{2n}=\frac{\pt-{2n}+\mt-{2n}}{2}\quad\hbox{and}\quad
{\bar t_{2n}}=\frac{\pt-{2n}-\mt-{2n}}{2}.\la{tot}
\eeq
Noting that \(\hamil[(1)]E=E\hamil[(1)]\) and
\(\hamil[(1)]{\bar E}=-{\bar E}\hamil[(1)]\), we have
\beqn
\Omega_{(0)}&\equiv& \evol[(1)]\left(
d\evol[(0)]{\evol[(0)]}^{-1}\right){\evol[(1)]}^{-1}\nonumber\\
&=&\left(1+\hamil[(1)]+\frac12{\hamil[(1)]}^2\right)
\left(dE+d{\bar E}\right)
\left(1-\hamil[(1)]+\frac12{\hamil[(1)]}^2\right)\nonumber\\
&=&dE+d{\bar E}\left(1-2\hamil[(1)]+2{\hamil[(1)]}^2\right).\la{omega0}
\eeqn
Taking into account the expressions \({\bar E}\) and
\(\hamil[(1)]^\pm\), we observe that
\[
(\der-\xi\xi-\xi\der-\xi)\hamil[(1)]^\pm=\pm\hamil[(1)]^\pm,
\qquad
(\der-\xi\xi-\xi\der-\xi)\hamil[(1)]^2=
[\hamil[(1)]^+,\;\hamil[(1)]^-]
\]
and hence
\beqn
(\der-\xi\xi-\xi\der-\xi)(\,\hamil[(1)]-{\hamil[(1)]}^2\,)
&=&\hamil[(1)]^+ - \hamil[(1)]^- - [\hamil[(1)]^+,\;\hamil[(1)]^-]
\nonumber\\
&=&\sum_{n=0}\left\{
\pt-{\podd-n}\,\left(\;\der-\xi+\frac12\sum_{m=0}
\mt-{\podd-m}{\der-x}^{n+m+1}\;\right) -
\left( \pt-{} \leftrightarrow \mt-{}\right)\right\}\nonumber\\
&=&\sum_{n=0}\left(
\pt-{\podd-n}\Omega^+_\podd-n - \mt-{\podd-n}\Omega^-_\podd-n\right).
\eeqn
Making use of the above equalities, we obtain the expressions
(\ref{evenP}) from (\ref{omega0}). As in the previous case,
assuming that \(\G\) is preserved through the time evolution
raised by \(\evol[M]\), we are led to Eq.(\ref{grasseq}).
Then the {\spdo} \(S(t)\) satisfies Eq.(\ref{skp}) with \(\Omega_M\).
{}From (\ref{oddP}) and (\ref{evenP}), Eq.(\ref{skp}) reads
\beqn
D^+_{\podd-n}S&=&-(S\der-\xi{\der-x}^nS^{-1})_-S,\nonumber\\
D^-_{\podd-n}S&=&-(S\xi{\der-x}^{n+1}S^{-1})_-S,\nonumber\\
D^+_{2n}S&=&-(S\der-\xi\xi{\der-x}^nS^{-1})_-S,\nonumber\\
D^-_{2n}S&=&-(S\xi\der-\xi{\der-x}^nS^{-1})_-S,\la{maxSKP}
\eeqn
where
\beqn
D^\pm_{\podd-n}&=&\der-{\pmt-{\podd-n}}-\frac12
\sum_{m=0}\mpt-{\podd-m}(\,\der-{\pt-{2(n+m+1)}}+\der-{\mt-{2(n+m+1)}}\,)
\nonumber\\
&=&\der-{\pmt-{\podd-n}}-\frac12
\sum_{m=0}\mpt-{\podd-m}\der-{t_{2(n+m+1)}},\la{oppmdd}\\
D^\pm_{2n}&=&\der-{\pmt-{2n}}\pm\sum_{m=0}
(\;\pt-{\podd-m}\der-{\pt-{\podd-{(n+m)}}}-
\mt-{\podd-m}\der-{\mt-{\podd-{(n+m)}}}\;).\la{oppmde}
\eeqn
The above system of equations defines the maximal SKP hierarchy.
It includes all the flows of MRSKP and JSKP as subflows.
This is first main result of the present paper.
Note that \(D^+_{2n}+D^-_{2n}=\der-{\pt-{2n}}+\der-{\mt-{2n}}
=\der-{t_{2n}}\) with \(t_{2n}\) being defined in (\ref{tot}).
The time differential operators \(D^\pm_n,\>\;(n\geq 1)\) satisfy
the commutation relations
\beqn
\{D^+_{\podd-m},\,D^-_{\podd-n}\}&=&-(D^+_{2(m+n+1)}+D^-_{2(m+n+1)})
=-\der-{t_{2(m+n+1)}},\nonumber\\
\{D^+_{\podd-m},\,D^+_{\podd-n}\}&=&\{D^-_{\podd-m},\,D^-_{\podd-n}\}=0,
\nonumber\\
{}[ D^+_{\podd-m},\,D^\pm_{2n}] &=&\pm D^+_{\podd-{(n+m)}},\nonumber\\
{}[ D^-_{\podd-m},\,D^\pm_{2n}]&=&\mp D^-_{\podd-{(n+m)}},\nonumber\\
{}[ D^+_{2m},\,D^\pm_{2n}]&=&[D^-_{2m},\,D^\pm_{2n}]=0.
\la{commuD}
\eeqn
This set of operators gives a representation of the Lie superalgebra
generated by the super vector fields,\\
\beq
\left\{\begin{array}{rl}
V^+_{\modd-n}&=\der-\xi{\der-x}^{n-1},\\
V^-_{\modd-n}&=\xi{\der-x}^n,
\end{array}\right.\qquad
\left\{\begin{array}{rl}
V^+_{2n}&=\der-\xi\xi{\der-x}^n,\\
V^-_{2n}&=\xi\der-\xi{\der-x}^n,
\end{array}\right.
\eeq
\(n=1,2,\cdots\), which are considered as a positive frequency part of
the generators of \({\widehat {gl1|1}}\). It is not hard to see that
the series of equations \(D_n^\pm S=-(SV_n^\pm S^{-1})_- S\) leads to
\[
[D_m^\pm,\,D_n^{\pm,(\mp)}]_\pm
S=(S[V_m^\pm,\,V_n^{\pm,(\mp)}]_{\pm} S^{-1})_- S,
\]
and hence the integrability requires (\ref{commuD}),
irrespectively of the explicit forms of the operators \(D_n^\pm\).

\subsection{Solution of the Grassmann equation}
We return to the Grassmann equation (\ref{grasseq}) and try
to solve it. The problem is reduced to that of linear
algebra by making use of the matrix representation introduced
in the previous section. Eq.(\ref{grasseq}) takes the form in the matrix
representation \cite{ike-yama}\footnote{
In writing \(U(t)\), we let \(t\) express the time dependence
symbolically. So hereafter when concerning the maximal SKP hierarchy,
\(t\) stands for \(\{t^\pm_n\}\).}
\beqn
&&\sum_{j\leq 1}s_j(x,\xi)\psi[U(t)S_0^{-1}]_{j,m}=0,\quad
m=1,2,\cdots\nonumber\\
&&\quad\hbox{with}\quad S(t)=\sum_{j\leq 1}s_j(x,\xi)D^j,\quad s_0=1.
\la{grasseq1}
\eeqn
{}From Proposition~\ref{pro1}, the matrix \(\psi[U(t)S_0^{-1}]\)
takes the form
\[
\psi[U(t)S_0^{-1}]=e^{\xximat}\Phi(t)\;\Xi_0e^{-\xximat},
\]
where we write
\beq
\Xi_0=\Xi[S_0^{-1}] \quad\hbox{and}\quad \Phi(t)=\Xi[U(t)],
\eeq
for simplicity, and use this notation in the following.
Since \((e^{\xximat})_{\geq <}={\bf 0}\) and
\((e^{\xximat})_{<<}\) is invertible, Eq.(\ref{grasseq1}) is
rewritten as
\beqn
&&\sum_{j\leq 0}s_j(x,\xi)Z(x,\xi,\,t)_{j,-m}=0,\quad
m=1,2,\cdots\nonumber\\
&&\quad\hbox{with}\quad Z(x,\xi,\,t)=e^{\xximat}\Phi(t)\;\Xi_0.
\la{grasseq2}
\eeqn
To solve Eq.(\ref{grasseq1}) or (\ref{grasseq2}), let us remind
ourselves of some basic ingredients of linear superalgebra.
Let \(X=(X_{ij})_{i,j\in\bZ}\) be an even homogeneous element
of \(\matZZA\) i.e. \(|X_{ij}|=i+j\>\;\hbox{(mod 2)}\), and let
\(X_I=(X_{ij})_{i,j\in I\cap\bZ}\) be a sub-matrix of \(X\) specified
by a set of indices \(I\subset\bZ\).
Supermatrices are often written in the block form with respect to
the Grassmann parity. Let us associate with \(X_I\) a matrix \(\check X_I\)
\beq
{\check X_I}=(X_I^{ab})_{a,b=0,1}\>,\qquad
X_I^{ab}=(X_{ij})_{\>i\in \,\{a+2\bZ\}\cap I,\;j\in \, \{b+2\bZ\}\cap I}.
\eeq
When the matrix \(X_I^{00}\) (\(X_I^{11}\)) is invertible, we define a matrix
\({\widehat X_I}^{00}\) (\({\widehat X_I}^{11}\)) by
\beq
{\widehat X_I}^{00}=X_I^{00}-X_I^{01}(X_I^{11})^{-1}X_I^{10},\quad
{\widehat X_I}^{11}=X_I^{11}-X_I^{10}(X_I^{00})^{-1}X_I^{01}.
\la{hatmat}
\eeq
The superdeterminant of the matrix \(X_I\) (or \(\check X_I\)) is defined,
as usual, by
\beq
{\rm sdet}X_I=\frac{\det {\widehat X_I}^{00}}{\det X_I^{11}}
=\frac{\det X_I^{00}}{\det {\widehat X_I}^{11}}.
\eeq
Keeping Eq.(\ref{grasseq1}) in mind, let us study an equation with
a given matrix \(X_{\wleinf}\)
\beq
v_{\leinf}X_{\leltinf}={\bf 0},\la{lineq}
\eeq
where \(v_{\leinf}=(v_j)_{\;j\leq 0}\) is a semi-infinite
row vector with a restriction \(v_0=1\). If both of the matrices
\(X_{\wleinf}\) and \(X_{\wltinf}\) are regular (i.e.
\(\det X^{00}_{\wleinf}\neq 0\), \(\det X^{00}_{\wltinf}\neq 0\)
and \(\det X^{11}_{\wleinf}=\det X^{11}_{\wltinf}\neq 0\;\)),
a unique solution for \(v_{\leinf}\) is given by
\beq
v_j=\frac{{\rm sdet}X_{\wleinf}}{{\rm sdet}X_{\wltinf}}\;
({X_{\wleinf}}^{-1})_{0\,j}\;,\quad j=0,-1,-2,\cdots.
\la{lineqsol}
\eeq
It is evident that \(v_{\leinf}\) given above satisfies
Eq.(\ref{lineq}), and \(v_0=1\) results from
\[
({X_{\wleinf}}^{-1})_{00}=({\widehat X_{\wleinf}}^{00}{}^{-1})_{00}
=\frac{\det {\widehat X_{\wltinf}}^{00}}{\det {\widehat X_{\wleinf}}^{00}}
=\frac{{\rm sdet}X_{\wltinf}}{{\rm sdet}X_{\wleinf}},
\]
here the last equality holds because in our notation
\( X^{11}_{\wleinf}=X^{11}_{\wltinf}\). Of course the right hand side
of (\ref{lineqsol}) depends only on the part \(X_{\leltinf}\).
In (\ref{lineqsol}),
\({\rm sdet}X_{\wleinf}\cdot({X_{\wleinf}}^{-1})_{0\,j}\) is considered
as a \((0,j)\)-cofactor and is expressed in terms of superdeterminants
as shown in \cite{ike-yama}, affording the supersymmetric generalization
of Cramer's formula. However the expression (\ref{lineqsol}) is more
convenient for later use. Returning to (\ref{grasseq2}) we observe
that
\(\left.Z(x,\xi,\,t)\right|_{x=\xi=t=0}=\Xi_0\) and
\((\Xi_0)_{ij}=\delta_{ij}\), for \(j\geq i\), and thus we can
conclude that formally
\(Z_{\wleinf}\) and
\(Z_{\wltinf}\) are both regular.
Consequently, Eq.(\ref{grasseq2}) has a unique solution
\beq
s_j=\frac{{\rm sdet}Z_{\wleinf}}{{\rm sdet}Z_{\wltinf}}\;
({Z_{\wleinf}}^{-1})_{0\,j}\;,\quad j=0,-1,-2,\cdots.\la{grasssol}
\eeq

The concrete expressions of \( \psi[U(t)]\) for each SKP hierarchies
are calculated as
\beqn
\psi[\evol[MR](t)]&=&
\exp\sum_{l=1}\left(t_{\modd-l}J\Gamma^{\modd-l}+t_{2l}\Lambda^l\right),
\\
\psi[\evol[J](t)]&=&
\exp\sum_{l=1}\left\{t_{\modd-l}(\pJ-\xi\Gamma)\Gamma^{\modd-l}
+t_{2l}\Lambda^l\right\},
\\
\psi[\evol[M](t^\pm)]&=&
\exp\sum_{l=0}\left\{\pt-{\podd-l}(\pJ-\xi\Gamma)\Gamma^{\podd-l}+
\mt-{\podd-l}(-\mJ+\xi\Gamma)\Gamma^{\podd-l}\right\}\nonumber\\
&&\qquad \times \>
\exp\sum_{l=1}\left\{\pt-{2l}(\pJ-\xi J\Gamma)\Lambda^l+
\mt-{2l}(\mJ+\xi J\Gamma)\Lambda^l\right\},
\la{evolop}
\eeqn
Here we note \(\Phi(t)\equiv\Xi[U(t)]=\left.\psi[U(t)]\right|_{\xi=0}\).

\subsection{Super-wave function}
The notion of the super-wave function (or Baker-Akhiezer function)
arises naturally from Eq.(\ref{grasseq}).
Taking Proposition~\ref{pro3}~(ii) into account
Eq.(\ref{grasseq}) is rewritten as
\beq
<S(t)U(t)e^{\xxi},\,\rho (S_0^{-1})\zeta^m>=0,\qquad m=0,1,2,\cdots.
\eeq
Here
\beq
\wvf\equiv S(t)U(t)e^{\xxi}\la{wave}
\eeq
is the super-wave function of the SKP hierarchy.
Remember that
\[
v^{(m)}=\rho (S_0^{-1})\zeta^m
=\sum_j\zeta^j(F_0)_{jm},\quad m=0,1,2,\cdots,\quad\hbox{with}\quad
F_0=JK\Xi_0KJ
\]
are the basis vectors (\ref{base}) of the subspace
\(W_{S_0}=\sigma (S_0)\in\sgrass\).
Therefore in terms of the wave function \(w\) of the form (\ref{wave}),
Eq.(\ref{grasseq}) is equivalently represented as a statement
\beq
\wvf\in {W_{S_0}}^\perp.\la{waveeq}
\eeq
where \({W_{S_0}}^\perp\) is the subspace orthogonal to \(W_{S_0}\)
with respect to the pairing (\ref{pairing}).\\
{}From (\ref{skp}) and (\ref{omega}), the super-wave function
satisfies
\beq
dw=(S\Omega S^{-1})_+\,w\>.
\eeq
Let us denote the factor \(U(t)\,e^{\xxi}\) by \(\facA\). Consider
the maximal SKP hierarchy and define formal power series
\beqn
\pwpmte-z&=&\sum_{m=1}\pmt-{2m}\,z^m,\nonumber\\
\pwte-z&=&\frac{\pwpte-z+\pwmte-z}{2},\qquad
\pwbarte-z=\frac{\pwpte-z -\pwmte-z}{2},\\
\pwpmtd-z&=&\sum_{m=0}\pmt-{\podd-m}\;z^m\nonumber\\
\pwtd-z&=&\frac{\pwptd-z+\pwmtd-z}{2},\qquad
\pwbartd-z=\frac{\pwptd-z -\pwmtd-z}{2},\\
\pwxipmt-z&=&\pwpmtd-z\pm \xi.\la{pwxipmt}
\eeqn
Then we have
\beqn
\facAM&=&\evol[(1)]\evol[(0)]\,\exp(\xxi)\nonumber\\
&=&\exp\left\{\; \pwptd-z\der-\xi+z\pwmtd-z\xi \;\right\}
\exp\left\{\; \pwpte-z\der-\xi\xi+\pwmte-z\xi\der-\xi \;\right\}\,
\exp(\xxi)\nonumber\\
&=&\exp xz\cdot\left\{\>
\exp\left\{\; \pwpte-z-\frac12
z\left(\propwtd-z+2\xi\pwmtd-z\right) \;\right\}\right.
\nonumber\\
&&\qquad\qquad\quad\left.
+\;\exp\left\{\pwmte-z+\frac12 z\propwtd-z\right\}\,(\pwptd-z+\xi)\theta
\>\right\}\la{facAM}\\
&=&\exp\left\{\; xz+\pwpte-z +\xi z\pwbartd-z+\;\pwxipt-z
\left(\theta\exp(-2\pwbarte-z)-\frac12 z \pwximt-z\right) \;\right\},\nonumber
\eeqn
and
\beq
\waveM=S(\pmt-{})\,\facAM.\la{wfM}
\eeq
The wave functions for JSKP and MRSKP hierarchies\cite{rabin}
are obtained from
(\ref{facAM}) and (\ref{wfM}) by replacing the time variables
as in (\ref{redtime}).
For JSKP, putting
\(\pwptd-z\rightarrow\pwtd-z\), \(\pwmtd-z=0\),
\(\pwbartd-z\rightarrow\frac12\pwtd-z\) and
\(\pwpte-z=\pwmte-z=\pwte-z\), we have
\beq
\supwvf-J=S(t)\exp \left(\,\xxi+\pwte-z +\pwtd-z\,\theta\,\right),
\eeq
and for MRSKP, putting
\(\pwptd-z=\pwmtd-z=\pwtd-z\) and
\(\pwpte-z=\pwmte-z=\pwte-z\), we have
\beq
\supwvf-{MR}=S(t)\exp \left\{\,xz+\xi \left(\theta - z\pwtd-z\right)
+\pwte-z+\pwtd-z\,\theta\;\right\}.
\eeq
Here we make an additional remark. From (\ref{Dexp})
we notice that, for \(P=\op[p,(j;j)]\in\E\),
\beq
Pe^{\xxi}=\sum_{j,m}p_j\,(e^{\xximat})_{jm}\,\zeta^m,
\eeq
and hence for \(P\), \(Q\;\in\E\),
\beq
PQe^{\xxi}=\sum_{j,m}p_j\;
(\psi[Q]e^{\xximat})_{jm}\,\zeta^m
=\sum_{j,m}p_j\,(e^{\xximat}\Xi[Q])_{jm}\,\zeta^m.\la{prodfour}
\eeq
Thus the wave function \(w\) has the expression
\beq
\wvf=\sum_m w_m\zeta^m\quad\hbox{with}\quad
w_m=\sum_j s_j\,(e^{\xximat}\Phi(t)\,)_{jm}.\la{wvmat}
\eeq

\subsection{Dual wave function and the bilinear identity}
Let us introduce an operation \(\star\) for invertible
{\spdos} by
\beq
P^\star\equiv(P^{-1})^\ast=(P^\ast)^{-1}.\la{starop}
\eeq
For example, consider this operation on the set of even monic
{\spdos} denoted by \(\E_0^\times\). Then \(\star\) defines
an endomorphism of \(\E_0^\times\). We also define an operation
\(\star\) for invertible (even homogeneous) matrices
\(X\in\matZZA\) by
\beq
X^\star\equiv IK(X^{\rm st})^{-1}KI.\la{matstar}
\eeq
Then, from (\ref{matadj}) we have
\beq
\psi[P^\star]=\psi[P]^\star,\qquad P\in\E_0^\times.\la{star}
\eeq
Applying the operation \(\star\) to (\ref{grasseq}), we obtain
the equation for \(S^\star (t)\):
\beq
(\,S^\star (t)U^\star (t){S_0^\star}^{-1}\,)_-=0.\la{dgrasseq}
\eeq
As in the case of wave function, we can write this equation in the form
\beq
<\,S^\star (t)U^\star (t)\,e^{-(\xxi)},\>
\left.\rho({S_0^\star}^{-1})\,\zeta^m
\right|_{{z\rightarrow -z}\atop {\theta\rightarrow -\theta}}>
=0,\qquad m=0,1,2,\cdots.\la{deqfourier}
\eeq
Here, for later convenience we put, in the integral defining the pairing,
\(z\rightarrow -z\), \(\theta\rightarrow -\theta\) (i.e.,
\(\zeta^m\rightarrow
(-1)^{\frac{m(m+1)}{2}}\zeta^m=I_{mm}\zeta^m,\;\>m\in\bZ\) ).
The dual wave function is given by
\beq
\dwvf=S^\star (t)U^\star (t)\,e^{-(\xxi)}.\la{dwave}
\eeq
Let us calculate the factor
\({\widetilde A}\equiv U^\star (t)\,e^{-(\xxi)}\). For the maximal
SKP hierarchy, we have
\beqn
\lefteqn{{\widetilde A_M}
(x,\xi,\pt-{\podd-n},\mt-{\podd-n},\pt-{2n},\mt-{2n}\,;\zeta)
=\evol[(1)]^\star (t)\evol[(0)]^\star (t)\,
\exp\{-(\xxi)\}}\nonumber\\
&&=\exp\sum_{n=0}\left\{\,\pt-{\podd-n}\der-\xi(-\der-x)^n
-\mt-{\podd-n}\xi(-\der-x)^{(n+1)}\,\right\}\nonumber\\
&&\qquad\quad\times\>
\exp-\sum_{n=0}\left\{\,\pt-{2n}\xi\der-\xi(-\der-x)^n
+\mt-{2n}\der-\xi\xi(-\der-x)^n\,\right\}\exp\{-(\xxi)\}
\nonumber\\
&&=\exp\sum_{n=0}\left\{\,-\pt-{\podd-n}\der-\xi{\der-x}^n
+\mt-{\podd-n}\xi{\der-x}^{(n+1)}\,\right\}\nonumber\\
&&\qquad\quad\times\>
\left.
\exp-\sum_{n=0}\left\{\,\mt-{2n}\der-\xi\xi{\der-x}^n
+\pt-{2n}\xi\der-\xi{\der-x}^n\,\right\}\exp{(\xxi)}
\right|_{{x\rightarrow -x}\atop {\xi\rightarrow -\xi}}
\nonumber\\
&&=A_M
(-x,-\xi,-\pt-{\podd-n},\mt-{\podd-n},-\mt-{2n},-\pt-{2n}\,;\zeta)
\nonumber\\
&&=\exp\left\{\;-xz-\pwmte-z -\xi z\pwbartd-z -\;\pwxipt-z
\left(\theta\exp(-2\pwbarte-z)-\frac12 z \pwximt-z\right) \;\right\}
\nonumber\\
&&=\exp 2\pwbarte-z \,\cdot {\facAM}^{-1}.\la{dfacAM}
\eeqn
In (\ref{deqfourier}) we see
\beq
{\widetilde v}^{(m)}\equiv
\left.(-1)^{\frac{m(m+1)}{2}}\rho({S_0^\star}^{-1})\,\zeta^m
\right|_{{z\rightarrow -z}\atop {\theta\rightarrow -\theta}}
=\sum_j\zeta^j (\dualframe)_{jm}\quad\hbox{with}\quad
\dualframe=IJK{\Xi_0}^\star KJI.
\eeq
Then from (\ref{matstar}),
\beq
\dualframe=J({\Xi_0}^{\rm st})^{-1}J.
\eeq
\(\dualframe\) is a triangular matrix
\((\dualframe)_{ij}=\delta_{ij}\), for \(j\geq m\), inheriting
the triangularity from \(\Xi_0=\Xi[S_0^{-1}]\). Therefore the
matrix \((\dualframe)_{\geinf}\) can be considered as a frame matrix
of a subspace in \(\H\), which we denote by \({\widetilde W_{S_0}}\),
defining an element of the {\sG} \(\sgrass\). Then
Eq.(\ref{deqfourier}) asserts
\beq
\dwvf\in {\widetilde W_{S_0}}^\perp.\la{dwaveeq}
\eeq
\indent
The validity of the bilinear identity for the (MR)SKP hierarchy has
been established in \cite{ike-yama}. Actually its validity does not
depend on the explicit forms of the time evolution operators.  For
completeness we state the bilinear identity with its proof.  Consider
the SKP hierarchy whose time evolution operator is \(U(t)\).
\begin{th}
Let \(\wvf=P\facA\) and \(\dwvf=Q\dfacA\),
\(P\), \(Q\in\sgrass\), where \(A=Ue^{\xxi}\) and
\({\widetilde A}=U^\star e^{-(\xxi)}\).
The necessary and sufficient condition for
\(w\) and \({\widetilde w}\) being the wave function and its dual
for the SKP hierarchy concerned is that they satisfy the following
bilinear identity:
\beq
\oint dzd\theta\;\wvf {\widetilde w}(x',\xi',t';\,\zeta)=0.\la{bilin}
\eeq
\end{th}
[{\it Proof}]\quad(Necessity).
Let \(P=Q^\star=S\) and set \(v^{(m)}=\sum_j\zeta^j(F_0)_{jm}\) and
\({\widetilde v}^{(m)}=\sum_j\zeta^j(\dualframe)_{jm}\) as before.
In addition to the fact that
\(\{v^{(m)}\}_{m\in\bZzp}\) and
\(\{{\widetilde v}^{(m)}\}_{m\in\bZzp}\) are the basises of
\(W_{S_0}\) and \({\widetilde W_{S_0}}\) respectively,
\(\{v^{(m)}\}_{m\in\bZ}\) and \(\{{\widetilde v}^{(m)}\}_{m\in\bZ}\)
are both considered as bases of \(\H\)
since \(F_0\) and \(\dualframe\) are both triangular
matrices with unit diagonal elements. Then
\beq
<\,{\widetilde v}^{(j)},\;v^{(m)}\,>=
(J\dualframe^{\rm st}JKF_0)_{jm}=K_{jm}=\delta_{j+m+1,0}\>\;,
\eeq
which says
\beq
W_{S_0}^\perp={\widetilde W_{S_0}}\quad\hbox{and}\quad
{\widetilde W_{S_0}}^\perp=W_{S_0}.
\eeq
Therefore, from (\ref{waveeq}) and (\ref{dwaveeq}), we find the
bilinear identity.\\
(Sufficiency). From Proposition~\ref{pro3}~(iii), we have
\((P(t)U(t)U(t')^{-1}Q(t')^\ast)_-=0\). Putting \(t_n=t'_n\)
and noting \(P,\>Q\in \sgrass\), we see \(Q=P^\star\). Moreover
from the bilinear identity,
\begin{eqnarray*}
0&=&\oint dzd\theta\;\der-{t_n}\wvf \;
{\widetilde w}(x',\xi',t';\,\zeta)\\
&=&\oint dzd\theta\;(\,B_n\wvf\,)
{\widetilde w}(x',\xi',t';\,\zeta),
\end{eqnarray*}
where \(B_n=(\der-{t_n}P)P^{-1}+P\Omega_n P^{-1}\).
Again from Proposition~\ref{pro3}~(iii), we have
\[
(\,(B_nPU)(t)(U^{-1}Q^\ast)(t')\,)_-=0.
\]
Putting \(t_n=t'_n\) together with the fact \(Q=P^\star\), we see
\((B_n)_-=0\), from which follows the Sato equations
\(\der-{t_n}P=-(P\Omega_nP^{-1})_-P^{-1}\).{}\hfill$\Box$\\\indent
Let us make a few remarks on the dual wave function.
\begin{enumerate}
\item
Let
\beq
S^\star(t)=\sum_{j\leq 0}(-1)^{\frac{j(j+1)}{2}}
{\widetilde s_j}\,D^j.
\eeq
{}From (\ref{dwave}) and taking account of (\ref{prodfour}),
we see the dual wave function has an expression
\beqn
\dwvf&=&\sum_{j,m}{\widetilde s_j}\,
(\;e^{-(x\Lambda+\xi J\Gamma)}I\Phi(t)^\star I\;)_{jm}\,\zeta^m
\nonumber\\
&=&\sum_{j,m}(-1)^m\zeta^m\,
(\;e^{-(x\Lambda+\xi J\Gamma)}I\Phi(t)^\star I\;)^{\rm st}_{mj}
{\widetilde s_j}\nonumber\\
&=&\sum_{j,m}\zeta^{-(m+1)}\,
(\;Je^{-(\xximat)}\Phi(t)^{-1}J\;)_{mj}
(J{\widetilde s})_{-(j+1)}.\la{dwvmat}
\eeqn
To obtain the above expression, we use the identities
\((\xi\Gamma)^{\rm st}=K\xi J\Gamma K\),
\(\Lambda^{\rm st}=K\Lambda K\),
\(I\Gamma I=-J\Gamma\) and
\(I\Lambda I=-\Lambda\).
\item
As the counterpart of Eq.(\ref{grasseq2}), rewriting
(\ref{dgrasseq}) as before, we have
\beq
\sum_{j\leq 0}(IZ^\star I)^{\rm st}_{mj}\,{\widetilde s_j}=0,
\quad m=-1,-2,\cdots,\quad\hbox{with}\quad
 Z=e^{\xximat}\Phi(t)\;\Xi_0.\la{dgrasseq2}
\eeq
We can solve this equation in terms of the matrix \(Z\)
as follows.
Noting \(K^{\rm st}=JK\) (\(K\) is a parity odd matrix),
we see from the definition
of the \(\star\) operation
\beq
(IZ^\star I)^{\rm st}KZ=-K,
\eeq
and hence especially
\beq
\sum_{j\in\bZ}(IZ^\star I)^{\rm st}_{mj}
Z_{-(j+1),k}=0,\quad\hbox{for}\>\;\hbox{\(m<0\) and \(k<0\)}.
\la{idzstz}
\eeq
Let us consider the \({\bf N}\times {\bf N}\) matrix
\beq
Z^{(j)}_{\wltinf}\equiv
\left(\begin{array}{c}
Z_{{}^{j<}}\\
Z_{{}^{\ll \!<}}
\end{array}\right),
\eeq
where the symbol \(\ll\) represents indices running on
\(\{-2,-3,\cdots\}\). In other words, \(Z^{(j)}_{\wltinf}\)
is a matrix \(Z_{\wltinf}\) except its \(-1\)st row is replaced
with the \(j\)-th row of the matrix
\(Z_{\ltinf}=\left(\begin{array}{c}
Z_{{}^{\geq <}}\\
Z_{{}^{<<}}
\end{array}\right)\).
Then from (\ref{idzstz}) we have
\beqn
0&=& \sum_{j\in\bZ}(IZ^\star I)^{\rm st}_{mj}\cdot
{\rm sdet}_{\pi}Z^{(-j-1)}_{\wltinf} \nonumber \\
 &=& \sum_{j\leq 0}(IZ^\star I)^{\rm st}_{mj}\cdot
{\rm sdet}_{\pi}Z^{(-j-1)}_{\wltinf}\> ,\label{idsdetz}
\eeqn
where \({\rm sdet}_{\pi}\) is defined, using the previous notations, as
\beq
{\rm sdet}_{\pi}X_I=\frac{\det {\widehat X_I}^{11}}{\det X_I^{00}}
\>\>(\;=({\rm sdet}X_I)^{-1}\;).
\eeq
The second equality in (\ref{idsdetz}) follows because
of the known property of the (super) determinant:
\[
{\rm sdet}_{\pi}
\left(\begin{array}{c}
Z_{{}^{j<}}\\
Z_{{}^{\ll \!<}}
\end{array}\right)=0,\quad\hbox{for}\>\;
j\in\{-2,-3,\cdots\}.
\]
Now from (\ref{idsdetz}) we have a solution of (\ref{dgrasseq2})
normalized by \({\widetilde s_0}=1\):
\beq
{\widetilde s_{-j}}=\frac{1}{{\rm sdet}_{\pi}Z_{\wltinf}}\;
{\rm sdet}_{\pi}Z^{(j-1)}_{\wltinf}\>,\qquad
j=0,1,2,\cdots.\la{grassdsol}
\eeq
\end{enumerate}

\resection{The operator theory for the SKP hierarchies}
In this section we reformulate the results in the preceding
sections in the field theoretic language.
\subsection{B-C system}
The B-C system\cite{fms} consists of a pair of superfields
\(B\) and \(C\) that take the forms
\beq
B(z,\,\theta)=\beta(z)+\theta b(z),\qquad
C(z,\,\theta)=c(z)+\gamma(z)\theta.
\eeq
We set, for the time being, the superconformal weights of \(B\) and \(C\)
to be \(0\) and \(\frac12\), respectively.
The mode expansions of the fields are
\beqn
B(z,\,\theta)&=&\sum_{j\in\bZ}(\bep-j+\theta b_j)z^{-j-1}
=\sum_{j\in\bZ}\zeta^{-j-1}B_j,\nonumber\\
C(z,\,\theta)&=&\sum_{j\in\bZ}(c_j+\gamm-j\theta)z^{-j}
=\sum_{j\in\bZ}C_j\zeta^{-j}.
\eeqn
Here we set
\beq
\left\{\begin{array}{lll}
B_{2j}&=&b_j,\\
B_{\podd-j}&=&\bep-j,
\end{array}\right.
\quad\hbox{and}\quad
\left\{\begin{array}{lll}
C_{2j}&=&c_j,\\
C_{\podd-j}&=&\gamp-j.
\end{array}\right.
\eeq
The oscillators satisfy the (anti-)commutation relations
\beqn
[C_m,\; B_n ]_\pm&=&C_mB_n-(-1)^{mn}B_nC_m=\delta_{m+n,0}\nonumber,\\{}
[C_m,\; C_n ]_\pm&=&[B_m,\;B_n]_\pm=0,\la{commurel}
\eeqn
and generate a non-commutative algebra (super Weyl algebra)
denoted by \(\alBC\). \(\alBC\) is a tensor product of the
Clifford algebra  \(\albc\) generated by the oscillators
\(b_n\) and \(c_n\)
and the Weyl algebra \(\albega\) generated by the oscillators
\(\bep-n\) and \(\gamp-n\):
\beq
\alBC=\albc\otimes\albega.
\eeq
Let us define a set of current operators
\beqn
j^+(z)&=&\mathop{\lim}_{w\rightarrow z}
\{\;-b(z)c(w)+\frac{1}{z-w}\;\}
=\sum_{n\in\bZ}\pj-n z^{-n-1}\nonumber,\\
j^-(z)&=&\mathop{\lim}_{w\rightarrow z}
\{\;-\beta (z)\gamma (w)-\frac{1}{z-w}\;\}
=\sum_{n\in\bZ}\mj-n z^{-n-1},\nonumber\\
\psi^+(z)&=&c(z)\beta (z)
=\sum_{n\in\bZ}\ppsip-n z^{-n-1},\nonumber\\
\psi^-(z)&=&\gamma (z)b(z)
=\sum_{n\in\bZ}\mpsip-n z^{-n-2}.\la{current}
\eeqn
These currents form generators of a super Lie algebra \(\gl11\).
We denote its envelope (i.e. super Heisenberg algebra) by \(\aljpsi\):
\beqn
[\pmj-m,\;\pmj-n]&=&\pm m\,\delta_{m+n,0},\nonumber\\{}
[\pj-m,\;\mj-n]&=&0,\\{}
[\pmj-m,\;\ppsi-r]&=&\pm\ppsi-{m+r},\nonumber\\{}
[\pmj-m,\;\mpsi-r]&=&\mp\mpsi-{m+r},\\
\{\ppsi-r,\;\mpsi-s\}&=&(\phalf-{-r})\,\delta_{r+s,0}-\tj-{r+s},
\quad\hbox{with}\quad \tj-m=\pj-m+\mj-m,\nonumber\\
\{\pmpsi-r,\;\pmpsi-s\}&=&0.
\eeqn
Note that the total current of the system
\(
j^t(z)\equiv j^+(z)+j^-(z)
\)
satisfies
\beq
[j^t(z),\;j^t(w)]=0\quad\hbox{and}\quad
[j^t(z),\;\psi^\pm (w)]=0,
\eeq
and especially \(\tj-0\) is a central element of \(\aljpsi\).\\
Next, we recall the Fock representations of the B-C system.
Let us define the vacuum state
\(|\ketcurr-0=|\ketbc-0 \otimes |\ketbega-0\) by
\beq
C_m|\ketcurr-0=0,\quad\hbox{for}\>\;m > 0,\quad\hbox{and}\quad
B_m|\ketcurr-0=0,\quad\hbox{for}\>\;m\geq 0
\eeq
and consider the Fock space (i.e. left \(\alBC\)-module)
generated from this state. We denote it by \(\F^{(0)}_{BC}\).
One can also consider the charged vacuum states
\beq
|\ketpq[p,q]=|\ketbc-p \otimes |\ketbega-q \>,\qquad p,q\in\bZ,
\eeq
where the states \(|\ketbc-p\) and \(|\ketbega-q\) are
specified by
\beq
\left.\begin{array}{lll}
\hbox{for}&n>-p,&c_n\\
\hbox{for}&n\geq p,&b_n
\end{array}\right\}
|\ketbc-p=0
\quad\hbox{and}\quad
\left.\begin{array}{lll}
\hbox{for}&r>p, &\gamma_r\\
\hbox{for}&r>-p,&\beta_r
\end{array}\right\}
|\ketbega-p=0.
\eeq
These state have expressions such as
\beqn
|\ketbc-p &=&\left\{\begin{array}{ll}
c_{-(p-1)}\cdots c_{-1}c_0|\ketbc-0\quad&\hbox{for}\>\; p>0,\\
b_p\cdots b_{-2}b_{-1}|\ketbc-0\quad&\hbox{for}\>\; p<0
\end{array}\right.\\
|\ketbega-p &=&\left\{\begin{array}{ll}
\delta(\bep-{-p})\cdots
\delta(\beta_{-\frac32})\delta(\beta_{-\frac12})
|\ketbega-0\quad&\hbox{for}\>\; p>0,\\
\delta(\gamp-p)\cdots
\delta(\gamma_{-\frac32})\delta(\gamma_{-\frac12})
|\ketbega-0\quad&\hbox{for}\>\; p<0.
\end{array}\right.
\eeqn
The regularization of the current \(j^\pm\)
appeared in (\ref{current}) are defined so that
\(\pj-0|\ketbc-0 =0\) and \(\mj-0|\ketbega-0 =0\),
and then
the state \(|\ketpq[p,q]\) possesses the following properties
for the current algebra \(\aljpsi\):
\beqn
(\pj-n - p\delta_{n0})|\ketpq[p,q]&=&0,
\qquad\hbox{for}\quad n\geq 0,\nonumber\\
(\mj-n - q\delta_{n0})|\ketpq[p,q]&=&0,
\qquad\hbox{for}\quad n\geq 0,\nonumber\\
\ppsi-r|\ketpq[p,q]&=&0,
\qquad\hbox{for}\quad r>-(p+q),\nonumber\\
\mpsi-r|\ketpq[p,q]&=&0,
\qquad\hbox{for}\quad r>p+q-1.\la{currannih}
\eeqn
Let us consider the Fock representation of \(\alBC\)
generated from \(|\ketpq[p,q]\).
Since
\[
\albc|\ketbc-p=\albc|\ketbc-0,
\] we have
a unique b-c Fock representation \(\F_{bc}\equiv\albc|\ketbc-0\).
On the other hand, since
\[
\albega |\ketbega-p\not=\albega|\ketbega-q
\quad\hbox{for}\>\;p\not= q,
\]
we have inequivalent Fock representations
\(\F^{(p)}_{\beta\gamma}=\albega|\ketbega-p\),
for every integer \(p\) called ``picture charge''\cite{fms}.
Therefore the Fock representations of the total B-C system
are given by
\beq
\F^{(p)}_{BC}\equiv \F_{bc}\otimes\F^{(p)}_{\beta\gamma},
\quad p\in\bZ.
\eeq
Actually, to realize the {\sG} within the language of the B-C
system, it suffices for us to work on the 0-th picture
\(\F^{(0)}_{BC}\).\\
To obtain correlation functions, we need to prepare bra-states.
The vacuum bra-states are defined through the involution \(\dagger\)
in \(\alBC\):
\beqn
b_n^{\dagger} &=&b_{-n},\qquad \gamma_r^{\dagger}=\gamma_{-r},\nonumber\\
c_n^{\dagger} &=&c_{-n},\qquad \beta_r^{\dagger}=-\beta_{-r}.
\eeqn
Then we see that
\beq
(\pmj-n)^{\dagger}=-\pmj-{-n}-Q^\pm\delta_{n,0},\qquad
{\pmpsi-r}^{\dagger}=\pmpsi-{-r}\>,
\eeq
where \(Q^\pm\) are the so called background charges which in our case
are given by
\beq
Q^+=-1\qquad\hbox{and}\qquad Q^-=0.
\eeq
We set
\beq
\brapq[p,q]|=(\,|\ketpq[p,q]\,)^{\dagger}.
\eeq
Then we have
\beq
\brabc-p |\,\pj-0=-(p-1)\>\brabc-p |\qquad\hbox{and}\qquad
\brabega-q |\,\mj-0=-q\>\brabega-q |.
\eeq
The relative normalizations of the bra- and ket-vacuum states
are fixed by
\beq
\brabc-{-p+1}|\ketbc-p=1\qquad\hbox{and}\qquad
\brabega-{-q}|\ketbega-q=1.
\eeq

\subsection{Representation of solutions for the Grassmann equation}
We know that the Fock space of the theory of free fermion such as
the b-c system is constructed in terms of semi-infinite wedge products
(or semi-infinite products of Grassmann odd variables).
We can extend this view in a supersymmetric way.
Let \(\tilcapc-{2n}=\tilc-n\) and
\(\tilcapc-{\podd-n}=\tilgamp-n,\>\;n\in\bZ\) be infinitely
many odd and even variables, respectively.
The algebra \(\alBC\) is represented through
\beq
C_j\Rightarrow\tilcapc-j\qquad\hbox{and}\qquad
B_j\Rightarrow (-1)^j\der-{\tilcapc-{-j}},
\eeq
and then the vacuum states are expressed as semi-infinite products
of delta functions:
\beq
|\ketpq[p,q]=\prod_{j=1-p}^{\infty}\tilc-j\,
\prod_{k=q}^{\infty}\delta (\tilgamp-k).
\eeq
Hereafter we use the notation
\(\prod_{j=p}^{\infty}\tilc-j\)
on the understandings that the indices of the odd variables
are increasing from left to right in its products:
\[
\prod_{j=p}^{\infty}\tilc-j=\tilc-p\tilc-{p+1}\tilc-{p+2}\cdots.
\]
Note that for odd variables \(\tilc-n=\delta (\tilc-n)\).\\
One can observe easily that the correlation functions are calculated
formally by multiple integrations:
\beqn
\lefteqn{
\brapq[p',q']|\,f(C_i,\;B_j)\,|\ketpq[p,q]}\nonumber\\
&=&\left.\int\prod_{r>-q'}d\tilgam-r
\int\prod_{l\geq p'}d\tilc-l\;f(\tilcapc-i,\;(-1)^j\der-{\tilcapc-{-j}})\;
\prod_{j\geq 1-p}\tilc-j\,
\prod_{s>q}\delta (\tilgam-s)
\right|_{{\tilgam-r=0,\>\;r<-q',}\atop{\tilc-l=0,\>\>\;l<p'\>.}}
\la{correl}
\eeqn
where \(f(C_i,\;B_j)\) is some polynomial function under consideration
and the integration over the odd variables
\(\int\prod_{j\geq p}d\tilc-j\) is Berezin integral defined as
\beq
\int\prod_{j\geq p}d\tilc-j\;h(\tilc-j)=
\left(\cdots\der-{\tilc-{p+2}}\der-{\tilc-{p+1}}\der-{\tilc-p}\,
h(\tilc-j)\,\right).
\eeq
We notice that the expression (\ref{correl}) is well defined
only when \(q'=-q\).\\
\indent
Now we link the B-C system with the SKP
hierarchy. Our argument here is straightforward.
Let us return to Eq.(\ref{grasseq2}) and its solution
(\ref{grasssol}). Consider a linear transformation
\beq
\tilcapc-j\mathop{\longrightarrow}^X
\tilcapc-{}(X)_j=\sum_k \tilcapc-k X_{-k,-j}\>,\quad j\in\bZ.
\la{lintransf}
\eeq
We associate the matrix \(X_{\ltinf}\) with a state
\beq
|X_{\ltinf}\rangle =
\prod_{j\geq 1}\tilc-{}(X)_j\prod_{s>0}\delta(\tilgam-{}(X)_s).
\eeq
Then, using the formula (\ref{correl}), we can show that
\beq
\brapq[1,0]|X_{\ltinf}\rangle=\brapq[0,0]|C_0|X_{\ltinf}\rangle
={\rm sdet}X_{\wltinf},\la{tauX}
\eeq
and moreover
\beq
\brapq[0,0]|C_j|X_{\ltinf}\rangle=\left\{
\begin{array}{cl}
{\rm sdet}X_{\wleinf}\cdot ({X_{\wleinf}}^{-1})_{0,-j}
\quad &\hbox{for}\quad j\geq 0,\\
&\\
0\quad &\hbox{for}\quad j<0\>.\end{array}\right.\la{sdetC}
\eeq
Thus the solution (\ref{grasssol}) can be expressed as
\beq
s_{-j}(x,\xi,\,t)=\frac{1}{\brapq[1,0]\,|\,Z_{\ltinf}(x,\xi,\,t)\rangle}\;
\brapq[0,0]|\,C_j\,|Z_{\ltinf}(x,\xi,\,t)\rangle,
\qquad j=0,1,2\cdots .\la{sol}
\eeq
One way to compute \(\brapq[0,0]|C_j|X_{\ltinf}\rangle\)
is to change the integration variables:
\beq
\tilcapc-j\longrightarrow \tilcapc-j'=
\sum_{k\geq 0} \tilcapc-k X_{-k,-j} +
\sum_{k<0} \tilcapc-k X_{-k,-j}=\tilcapc-{}(X)_j,\quad j=0,1,2\cdots ,
\la{ch-var}
\eeq
where \(\tilcapc-{\ltinf}\) are considered as auxiliary constants.
Then the Jacobian is given by
\beq
\int\prod_{r>0}d\tilgam-r
\int\prod_{j\geq 0}d\tilc-j={\rm sdet}X_{\wleinf}\;
\int\prod_{r>0}d\tilgam-r'\int\prod_{j\geq 0}d\tilc-j'\>.
\eeq
Expressing \(\tilcapc-j,\>\;j\geq0\) in the integrand
in terms of \(\tilcapc-{\geinf}'\) and \(\tilcapc-{\ltinf}\),
we have for \(j\geq 0\),
\beqn
\lefteqn{
\brapq[0,0]|C_j|X_{\ltinf}\rangle=
\int\prod_{r>0}d\tilgam-r\int\prod_{k\geq 0}d\tilc-k\>
\tilcapc-j\>
\left.\prod_{k\geq 1}\tilc-{}(X)_k\prod_{r>0}\delta(\tilgam-{}(X)_r)
\right|_{\tilcapc-{\ltinf}=0}}\nonumber\\
&=&{\rm sdet}X_{\wleinf}\;
\int\prod_{r>0}d\tilgam-r'\int\prod_{k\geq 1}d\tilc-k'
\int d\tilc-0'
\nonumber\\&&\qquad\qquad\quad
\sum_{k\geq0}(\,\tilcapc-k' - \sum_{l>0}\tilcapc-{-l}X_{l,-k}\,)
({X_{\wleinf}}^{-1})_{-k,-j}\;
\left.\prod_{k\geq 1}\tilc-k'\prod_{r>0}\delta(\tilgam-r')
\right|_{\tilcapc-{\ltinf}=0}\nonumber\\
&=&{\rm sdet}X_{\wleinf}\cdot ({X_{\wleinf}}^{-1})_{0,-j}\>.
\la{calsdetC}
\eeqn
Here we take the convention as if the factor
\(\int\prod_{j\geq 1}d\tilc-j\) is Grassmann even:
\beq
\int\prod_{j\geq 1}d\tilc-j\>\lambda
=\lambda\,\int\prod_{j\geq 1}d\tilc-j\quad
\hbox{for an arbitrary odd const. \(\lambda\)},\la{conven}
\eeq
which is equivalent to the requirement
\beq
\brapq[1,0]|\lambda=\lambda\brapq[1,0]|\quad\hbox{and}\quad
\lambda|\ketpq[0,0]=|\ketpq[0,0]\lambda.
\eeq
We can also show that the expression
(\ref{grassdsol}) for the solution of Eq.(\ref{dgrasseq2})
can be reproduced by
\beq
(-1)^j{\widetilde s_{-j}}(x,\xi,\,t)=
\frac{1}{\brapq[1,0]\,|\,Z_{\ltinf}(x,\xi,\,t)\rangle}\>
\brapq[1,0]|\,\gamma_{\frac12}B_{j-1}\,|Z_{\ltinf}(x,\xi,\,t)\rangle,
\quad j=0,1,2,\cdots .\la{dsol}
\eeq
The formula above can also be derived from a similar calculus as
in (\ref{calsdetC}), which seems of some interest, so
we present it in Appendix A.

\subsection{Lie algebra \(\liea (\ala)\)}
Let us define the superlinear space
\beq
\spaceH=(\mathop{\oplus}_{j\in\bZ}\complnum\,C_j)\otimes \ala.
\eeq
Of course there is correspondence between \(\spaceH\) and \(\H\)
with \(C_j\leftrightarrow \zeta^{j-1}\).
According to (\ref{lintransf}), consider the \(GL(\spaceH)\)
transformation
\beq
C(X)_j=\sum_{k\in\bZ}C_kX_{-k,-j}.\la{transfC}
\eeq
So far we do not be careful with the infinite dimensionality.
Here we set \(X\) belongs \(GL(\spaceH)\),
if and only if
\[X_{ij}=0,\qquad\hbox{for}\quad |i-j|\gg 1,\]
and \(X\) is invertible. The commutation relation
(\ref{commurel}) is then preserved provided the transformation
of the oscillators \(B_j\)
\beq
B(X)_j=(JX^{-1}J)_{jk}B_k \la{transfB}
\eeq
is accompanied. Because of this symmetry we can construct,
in the Fock space \(\F^{(0)}_{BC}\),
representations of \(GL(\spaceH)\) and the associated Lie algebra
\(\liea (\ala)\) defined below.
First we introduce the Lie superalgebra (over \(\complnum\))
\(\lieabar=
{\bar a}_{{\infty|\infty}_0}\oplus{\bar a}_{{\infty|\infty}_1}\) :
\beqn
\lieabar&=&\{\,X=(X_{ij})\in Mat(\bZ\times\bZ,\>\complnum)\,|
\; X_{ij}=0,\quad\hbox{for}\>\;|i-j|\gg 1\,\},\nonumber\\
{\bar a}_{{\infty|\infty}_0}
&=&\{\,X\in\lieabar\;|\;X^{01}=X^{10}={\bf 0}\,\},\nonumber\\
{\bar a}_{{\infty|\infty}_1}
&=&\{\,X\in\lieabar\;|\;X^{00}=X^{11}={\bf 0}\,\}.
\eeqn
Then the Lie algebra \(\lieabar (\ala)\) is defined as the even part
of the tensor product \(\lieabar\otimes\ala\)\cite{bergv}.
Now we associate the quadratic operators with
\(X\in\lieabar\otimes\ala\):
\beqn
J[X]&=&\sum_{j,k}\,:C_jX_{-j,k}B_k:(-1)^k
=\sum_{j\leq k}C_j X_{-j,k}B_k(-1)^k
+\sum_{j>k}B_k X_{-j,k}C_j(-1)^{j(j+k)}\nonumber\\
&=&\sum_{j,k}\left\{\,C_j X_{-j,k}B_k(-1)^k-
\brapq[1,0]|\;C_j X_{-j,k}B_k\;|\ketpq[0,0](-1)^k\,\right\}.
\eeqn
Then we see \(J[X]\) generates the infinitesimal transformation
of (\ref{transfC}) and (\ref{transfB}):
\beq
[J[X],\,C_j]=\sum_k C_kX_{-k,-j}\>,\qquad
[J[X],\,B_j]=-\sum_k (JXJ)_{jk}B_k.
\eeq
Commutation relations among the quadratic operators are
\beq
[\;J[X],\,J[Y]\;]=J[\,[X,\,Y]\,]+C(X,Y),
\eeq
where
\beq
C(X,Y)=\sum_{i\leq -1}\sum_{j\geq 0}
\{\,X_{ij}Y_{ji}(-1)^i-X_{ji}Y_{ij}(-1)^j\,\}.\la{center}
\eeq
Note that the right-hand side of (\ref{center}) is a finite sum
when \(X,Y\in\lieabar (\ala)\). The above commutation relation
defines a central extension of \(\lieabar (\ala)\) :
\beq
\liea (\ala)=\lieabar (\ala)\oplus \ala_0 1.\la{liea}
\eeq

\subsection{Formula for the (dual) super-wave function}
The oscillators (with positive frequencies) in the current operators
(\ref{current}) are expressed as
\beqn
\pj-n&=&\sum_k C_{2(k+n)}B_{-2k}=J[\frac{1+J}{2}\Lambda^n],
\nonumber\\
\mj-n&=&\sum_k -C_{2(k+n)+1}B_{-2k-1}=J[\frac{1-J}{2}\Lambda^n],
\nonumber\\
\lambda\,\ppsip-n
&=&\sum_k \lambda\,C_{2(k+n)}B_{-2k+1}
=J[\lambda\frac{1+J}{2}\Gamma^\podd-n],
\nonumber\\
\lambda\,\mpsip-n
&=&\sum_k \lambda\,C_{2(k+n)+1}B_{-2k}
=J[\lambda\frac{1-J}{2}\Gamma^\podd-n],
\eeqn
all for \(n\in\bZ_{\gtinf}\),
where \(\lambda\) is an odd parameter.
Therefore, remembering the time evolution matrix \(\Phi_M(t^\pm)\)
obtained from (\ref{evolop}) and using the formulas obtained in the
previous section, we have
\beqn
\sum_{j,m} C_{-j}(e^{\xximat}\Phi_M(t^\pm)\lefteqn{)_{jm}\,\zeta^m
=\sum_j C_{-j}D^j\facAM} \nonumber\\
&=&e^{\opxxi}\;\opevol_M(t^\pm)
\,C(\zeta)\,\opevol_M(t^\pm)^{-1}e^{-(\opxxi)},\nonumber\\
\sum_{j,m}\zeta^{-(m+1)}
(Je^{-(\xximat)}\Phi_M \lefteqn{(t^\pm)^{-1}J)_{mj}B_j
=\sum_j(-1)^{\frac{j(j+1)}{2}}B_{j-1}D^{-j}\dfacAM} \nonumber\\
&=&e^{\opxxi}\;\opevol_M(t^\pm)
\,B(\zeta)\,\opevol_M(t^\pm)^{-1}e^{-(\opxxi)}\la{factor},
\eeqn
where
\beq
\sigma=\ppsi-{\frac12} +\mpsi-{\frac12},
\qquad
\tj-1=\pj-1 +\mj-1,
\eeq
and
\beq
\opevol_M(t^\pm)=\exp(\opppsi -\opmpsi)\,\exp(\oppj +\opmj).\la{opevol}
\eeq
Here we use the following notations:
\beqn
\oppmj&\equiv&\sum_{n=1}\pmt-{2n}\,\pmj-n
=\frac{1}{2\pi i}\oint dz\,\pwpmte-z j^\pm(z),\\
\oppmpsi&\equiv&\sum_{n=0}\pmt-{\podd-n}\pmpsip-n=
\left\{\begin{array}{lcl}
\frac{1}{2\pi i}\oint dz\, \pwptd-z\psi^+(z)&\hbox{for}&\opppsi,\\
&&\\
\frac{1}{2\pi i}\oint dz\, z\pwmtd-z\psi^-(z)&\hbox{for}&\opmpsi.
\end{array}\right.
\eeqn
At this stage we are led to field theoretic expressions for the
wave function and the dual wave function from
(\ref{wvmat}), (\ref{dwvmat}), (\ref{sol}) and (\ref{dsol}), which
are the main results of this section:
\beqn
\wvf&=&\frac{1}{\tau\xxitz}\;
\brapq[0,0]|\,e^{\opxxi}\,\opevol(t)C(\zeta)\,
|\ketcurr-{{\Xi_0}_{\ltinf}},\la{Cwavef}\\
\dwvf&=&\frac{1}{\tau\xxitz}\;
\brapq[1,0]|\,\gamma_{\frac12}e^{\opxxi}\,\opevol(t)B(\zeta)\,
|\ketcurr-{{\Xi_0}_{\ltinf}},\la{Bwavef}
\eeqn
where
\beq
\tau\xxitz=\brapq[1,0]\,|\,Z_{\ltinf}(x,\xi,\,t)\rangle
=\brapq[1,0]|\,e^{\opxxi}\,\opevol(t)\,|\ketcurr-{{\Xi_0}_{\ltinf}}
\la{tauf}
\eeq
is considered as the tau function of the SKP hierarchy.
However, one needs \(\tauM(t^\pm)\) of the maximal case in order
to draw the wave function from the \(\tau\)-function.\\\indent
For a given \(X\in GL(\spaceH)\), we can find an element
\(\opO-X\) in \(\alBC\) such that
\beq
\opO-X C_j\opO-X^{-1}=C(X)_j\quad\hbox{and}\quad
\opO-X B_j\opO-X^{-1}=B(X)_j.
\eeq
The associated state \(|\ketcurr-{X_{\ltinf}}\) lies on
\(GL(\spaceH)\) orbit of the vacuum state, i.e.,
\(|\ketcurr-{X_{\ltinf}}=\opO-X|\ketpq[0,0]\), and is viewed
as a section of the dual determinant bundle \(DET^\ast\) over
the \(\G\)\cite{bergv}.
\\ \indent
Although we take here a direct way to show that
the wave function and its dual can be expressed in such forms of
(\ref{Cwavef}) and (\ref{Bwavef}), this can be simply derived from
(\ref{factor}) and the fact that (\ref{Cwavef}) and (\ref{Bwavef})
satisfy the bilinear identity (\ref{bilin}). Since
\(\oint dz d\theta\,C(\zeta)\otimes B(\zeta)\)
is \(GL(\spaceH)\) invariant, we have the super-Pl\"ucker
equation\cite{kac-vdleur}
\beq
\oint dz d\theta\,C(\zeta)\,|\ketcurr-{{\Xi_0}_{\ltinf}}
\otimes B(\zeta)\,|\ketcurr-{{\Xi_0}_{\ltinf}}=0,\la{BCbilin}
\eeq
from which (\ref{bilin}) immediately follows.\\
\indent
Let us mention about the characteristics of the known SKP
hierarchies which we can see immediately.
For JSKP hierarchy, the operator (\ref{opevol}) reduces
\beq
\opevol_J(t)=\exp(\psi^+[t_o]+\optj),
\eeq
and we see
\beq
\psi^+[t_o]+\optj=\frac{1}{2\pi i}\oint dz d\theta\,
\{\,\pwtd-z +\pwte-z \theta\,\}J(z,\,\theta),\la{hamil-jskp}
\eeq
where
\(J(\zeta)\) is the superconformal current
\beq
J(\zeta)=-:B(\zeta)C(\zeta):.
\eeq
Therefore the super-wave function (\ref{Cwavef}) in this case
respects the superconformal structure (i.e. can be viewed as
a superconformal field), if the state \(|\ketcurr-{{\Xi_0}_{\ltinf}}\)
is associated with a super-Riemann surface by the
super-Krichever map\cite{mula-rab}.
On the other hand for MRSKP hierarchy, we have
\beq
\opevol_{MR}(t)=\exp\{\,(\psi^+[t_o]+\optj)-\psi^-[t_o]\,\}.
\eeq
Here \(\psi^-(z)=b(z)\gamma (z)\) is one of the charged fermionic
generators of the \(N=2\) super Virasoro algebra of the B-C system.
Therefore deformation of the geometric data through \(\psi^-[t_o]\)
generates variation in the moduli of the supercurve.\cite{rabin} Thus
the superconformal structure will be violated even if it exists in
the initial data, in the context of the \(N=1\) superconformal
symmetry.

\resection{Tau function description of the SKP hierarchy}\label{sec:tau}
Our main goal in this section is to express the (dual) wave function
from the tau function through vertex operators.
To do this we recall, as an intermediate step, the superbosonization
scheme introduced by Kac and van de Leur\cite{kac-vdleur}.
Then we construct the vertex operators which act to the \(\tau\)-function
and give the effect of inserting B and C fields in the correlation
that represents tau function.

\subsection{Kac-van de Leur superbosonization}
In \cite{kac-vdleur} Kac-van de Leur established a super
boson-fermion correspondence between the Clifford (or Weyl)
superalgebra and the super Heisenberg algebra
\(\aljpsi\)\footnote{Actually, there is no essential difference
between the Clifford and Weyl case.}. We describe here their
construction heuristically, with notational change pushing
forwards that their Weyl superalgebra is just the B-C system.
Since the fermionic \(b\), \(c\) and bosonic \(\beta\) and \(\gamma\)
fields carry their own charges, we have to introduce, in addition
to the \(\gl11\) currents, the operators \(e^{\pmphi-0}\), with
\(\pmphi-0\) being the conjugate operators with \(\pmj-0\),
which produce the charge sector in the Fock spaces:
\beqn
[\,\pmj-0,\;e^{\pmphi-0}\,]&=&e^{\pmphi-0},\nonumber\\{}
[\,\pmj-0,\;e^{\mpphi-0}\,]&=&0
\quad\hbox{and}\quad
[\,\pmj-n,\;e^{\pphi-0}\,]=[\,\pmj-n,\;e^{\mphi-0}\,]=0
\quad\hbox{for}\>\;n\not= 0.
\eeqn
In addition \(e^{\pmphi-0}\) satisfy
\beq
e^{-\pphi-0}\pmpsi-r\, e^{\pphi-0}=\pmpsi-{r\pm 1},
\qquad
e^{-\mphi-0}\pmpsi-r\, e^{\mphi-0}=\pmpsi-{r\pm 1}.
\la{chgvtx}
\eeq
These relations are most easily recognized from the well known
Friedan-Marinec-Shenker (FMS)
bosonization formula of the B-C system\cite{fms}.
Let us introduce the fields
\beq
\phiani{\pm}(z)=\pm \sum_{n>0}\frac{-1}{n}\,\pmj-n\, z^{-n}
\quad\hbox{and}\quad
\phicri{\pm}(z)=\pm \sum_{n>0}\frac1n\,\pmj-{-n}\, z^n.
\eeq
As for the fermionic \(b\), \(c\) fields, the superbosonization
rule is nothing but the ordinary bosonization:
\beq
c(z)=e^{\pphi-0}z^{\pj-0}e^{\phicri+ (z)}e^{\phiani+ (z)},\qquad
b(z)=e^{-\pphi-0}z^{-\pj-0}e^{-\phicri+ (z)}e^{-\phiani+ (z)}.
\la{bfbc}
\eeq
A simple way to see the rules for the bosonic \(\beta\) and \(\gamma\)
fields is to make use of the following OPE's:
\beq
\psi^+(z)\,b(w)\sim \frac{1}{z-w}\,\beta (z),\qquad
\psi^-(z)\,c(w)\sim \frac{1}{z-w}\,\gamma (z).
\eeq
Putting the expression (\ref{bfbc}) and taking into account the
identities
\beqn
e^{\pm\phicri+ (w)}\,\psi^\pm(z)\,e^{\mp\phicri+ (w)}&=&
\frac{1}{1-\frac{w}{z}}\;\psi^\pm(z),\quad\hbox{for}\>\;|z|>|w|,\\
e^{\pm\pphi-0}\psi^\pm(z)\, e^{\mp\pphi-0}&=&\frac1z\;\psi^\pm(z),
\eeqn
one finds the following superbosonization rule.
\beqn
\gamma(z)&=&
e^{\pphi-0}z^{1+\pj-0}e^{\phicri+ (z)}\psi^-(z)\,e^{\phiani+ (z)},
\nonumber\\
\beta(z)&=&
e^{-\pphi-0}z^{1-\pj-0}e^{-\phicri+ (z)}\psi^+(z)\,e^{-\phiani+ (z)}.
\la{bfbega}
\eeqn
For our purpose it is enough to know the formulas (\ref{bfbc})
and (\ref{bfbega}). We notice that the current
\(j^-=-:\beta\gamma:\) (or \(\phi^-_{{}^>_<}\)) does not appear
explicitly in these formulas. In particular we do not need the
operator \(e^{\pm\mphi-0}\) (which changes the picture number),
because the charges of the \(\beta\), \(\gamma\) are carried through
\(\psi^\pm\) as well. In addition to \(\beta\) and \(\gamma\), there exist
fields which intermediate different Fock spaces (i.e.  different
pictures) \(\F^{(p)}_{\beta\gamma}\) as we have seen in the previous
section. So much for the superbosonization.
We defer the rest to Appendix~\ref{append; b}.
Next, let us consider the representation of the super-Heisenberg
algebra \(\aljpsi\) on the Fock space \(\F^{(0)}_{BC}\).
In \cite{kac-vdleur} it was shown that \(\F^{(0)}_{BC}\) is
decomposed into a direct sum of irreducible representations of
\(\aljpsi\) according to the total charge number \(m\in\bZ\):
\beq
\F^{(0)}_{BC}=\mathop{\oplus}_{m\in\bZ}\F^{(0)}_m,
\quad\hbox{with}\quad\tj-0\;\F^{(0)}_m=m\,\F^{(0)}_m.
\eeq
{}From (\ref{currannih}) we observe that the highest state
\(|\ketcurr-m\) which specifies the representation \(\F^{(0)}_m\)
is not \(|\ketpq[m,0]\) itself but is given by
\beq
|\ketcurr-m=\left\{\begin{array}{lcl}
\mpsi-{\frac12}\cdots \mpsi-{m-\frac32}\,|\ketpq[m,0]
=(\gamma_{-\frac12})^{m-1}|\ketpq[1,0],&\hbox{for}&m\geq1,\\
&&\\
\ppsi-{\frac12}\cdots \ppsim-{|m|}\,|\ketpq[m,0]
=(\beta_{-\frac12})^{|m|}|\ketpq[0,0],&\hbox{for}&m\leq 0.
\end{array}\right.
\la{chgvcm}
\eeq
Then the state \(|\ketcurr-m,\>\;m\in\bZ\) satisfies the highest
weight conditions
\beqn
\pj-0\;|\ketcurr-m &=&\left\{\begin{array}{ccl}
|\ketcurr-m,&\>\hbox{for}&m\geq1,\\
0,&\>\hbox{for}&m\leq 0,\end{array}\right.
\qquad
\mj-0\;|\ketcurr-m =\left\{\begin{array}{ccl}
(m-1)|\ketcurr-m,&\>\hbox{for}&m\geq1,\\
m\;|\ketcurr-m,&\>\hbox{for}&m\leq 0,\end{array}\right.
\nonumber\\
\pmj-n\;|\ketcurr-m &=&0,\quad\hbox{for}\quad n>0
\quad\hbox{and}\quad
\pmpsi-r\;|\ketcurr-m =0,\quad\hbox{for}\quad r>0.
\eeqn
The highest weight states for the right
\(\aljpsi\)-modules are similarly defined through
\(\bracurr-m|=( |\ketcurr-m)^\dagger\). We notice that the states
\(\bracurr-1|\) (\(|\ketcurr-1\)) and
\(\bracurr-0|\) (\(|\ketcurr-0\)) satisfy the additional conditions
respectively,
\beqn
\bracurr-1|\;\ppsi-{\frac12}&=&0
\quad (\ppsi-{-\frac12}\,|\ketcurr-1=0),\nonumber\\
\bracurr-0|\;\mpsi-{\frac12}&=&0
\quad (\mpsi-{-\frac12}\,|\ketcurr-0=0).
\eeqn
Note also that for an odd parameter \(\lambda\),
\beq
\bracurr-m|\,\lambda=\left\{\begin{array}{ccl}
\lambda\,\bracurr-m|&\hbox{for}&m\geq1,\\
-\lambda\,\bracurr-m|&\hbox{for}&m\leq 0,\end{array}\right.
\eeq
which follows from our convention (\ref{conven}) for the state
\(\bracurr-1|\) and from (\ref{chgvcm}).

\subsection{Vertex operators}
We are now in a position to construct the vertex operators which are
defined as
\beqn
\lefteqn{
V_C\zexxipmt\bracurr-1|\,e^{\opxxi}\opevol(t^\pm)=
\bracurr-0|\,e^{\opxxi}\,\opevol(t^\pm)\,C(\zeta)}\nonumber\\
&=&\bracurr-0|\{\,c(z)+\gamma(z)(\der-{\xi'}+\xi'\der-{x'})\,\}\,
e^{\opxxi}\,\opevol(t^\pm)
\left.A_M(x',\xi',\pmt-{}\,;\zeta)\right|_{x'=x,\>\,\xi'=\xi},
\la{vopC}\nonumber\\
\lefteqn{
V_B\zexxipmt\bracurr-1|\,e^{\opxxi}\opevol(t^\pm)=
\bracurr-2|\,e^{\opxxi}\,\opevol(t^\pm)\,B(\zeta)}\nonumber\\
&=&\bracurr-2|\{\,\beta(z)+b(z)(\der-{\xi'}+\xi'\der-{x'})\,\}\,
e^{\opxxi}\,\opevol(t^\pm)
\left.{\widetilde A_M}
(x',\xi',\pmt-{}\,;\zeta)\right|_{x'=x,\>\,\xi'=\xi},
\la{vopB}
\eeqn
where \(\opevol\), \(A_M\) and \({\widetilde A_M}\) are given by
(\ref{opevol}), (\ref{facAM})
and (\ref{dfacAM}), respectively.
Given these operators, we can express the wave function and its dual in
terms of \(\tau\) function as
\beqn
\waveM&=&\frac{1}{\tauM\xxipmtz}\,V_C\tauM\xxipmtz,\nonumber\\
\dwaveM&=&\frac{1}{\tauM\xxipmtz}\,V_B\tauM\xxipmtz.
\eeqn
In order to obtain the above vertex operators, it is sufficient to
find the operators
\beqn
v_c\zxxipmt\,\bracurr-1|\,e^{\opxxi}\opevol(t^\pm)&=&
\bracurr-0|\,c(z)\,e^{\opxxi}\opevol(t^\pm),\nonumber\\
v_\gamma\zxxipmt\,\bracurr-1|\,e^{\opxxi}\opevol(t^\pm)&=&
\bracurr-0|\,\gamma (z)\,e^{\opxxi}\opevol(t^\pm),\nonumber\\
v_b\zxxipmt\,\bracurr-1|\,e^{\opxxi}\opevol(t^\pm)&=&
\bracurr-2|\,b(z)\,e^{\opxxi}\opevol(t^\pm),\nonumber\\
v_\beta\zxxipmt\,\bracurr-1|\,e^{\opxxi}\opevol(t^\pm)&=&
\bracurr-2|\,\beta (z)\,e^{\opxxi}\opevol(t^\pm).
\eeqn
\(V_C\) and \(V_B\) are then given by
\beqn
V_C\zexxipmt&=&A_M\,v_c\zxxipmt
-(DA_M)\,v_\gamma\zxxipmt,\nonumber\\
V_B\zexxipmt&=&{\widetilde A_M}\,v_\beta\zxxipmt
-(D{\widetilde A_M})\,v_b\zxxipmt.
\eeqn
For simplicity, let us consider the case \(x=\xi=0\) and define
\beqn
v_c\zpmt\,\bracurr-1|\,\opevol(t^\pm)=
\bracurr-0|\,c(z)\opevol(t^\pm),&&\;
v_b\zpmt\,\bracurr-1|\,\opevol(t^\pm)=
\bracurr-2|\,b(z)\opevol(t^\pm),\nonumber\\
v_\gamma\zpmt\,\bracurr-1|\,\opevol(t^\pm)=
\bracurr-0|\,\gamma (z)\opevol(t^\pm),&&\;
v_\beta\zpmt\,\bracurr-1|\,\opevol(t^\pm)=
\bracurr-2|\,\beta (z)\opevol(t^\pm).
\eeqn
Provided we can find these operators, the expressions for the general
case with \(x\not= 0\) and \(\xi\not= 0\), can be found using the identity
\beqn\lefteqn{
e^{\opxxi}\,\opevol(\,\pwptd-{\bullet},\,\pwmtd-{\bullet},\,
\pwpte-{\bullet},\,\pwmte-{\bullet}\,)}\nonumber\\
&=&\opevol(\,\pwxipt-{\bullet},\;\pwximt-{\bullet},\;
\pwpte-{\bullet}+x\bullet+\xi\bullet\pwbartd-{\bullet},\;
\pwmte-{\bullet}+x\bullet+\xi\bullet\pwbartd-{\bullet}\,),\la{opxxi-x}
\eeqn
where \(\opevol\) is considered as a functional of
\(t^\pm_{o,e}(z)\) and \(\bullet\) represents a dummy variable.
Now putting the superbosonization formulas for \(c\), \(\gamma\),
\(b\) and \(\beta\), and noting
\begin{eqnarray*}
\bracurr-2|\,e^{-\pphi-0}z^{-\pj-0}e^{-\phicri+(z)}
&=&\bracurr-1|\,\gamma_{\frac12}e^{-\pphi-0}z^{-\pj-0}e^{-\phicri+(z)}\\
&=&z^{-1}\,\bracurr-0|\,\gamma_{\frac12}
=z^{-1}\,\bracurr-1|\,\mpsi-{\frac12},\\
\bracurr-2|\,e^{-\pphi-0}z^{-\pj-0}e^{-\phicri+(z)}\psi^+(z)
&=&z^{-1}\,\bracurr-1|\,\mpsi-{\frac12}\psi^+(z)
=z^{-1}\,\bracurr-1|\left(\{\mpsi-{\frac12},\,\ppsi-{-\frac12}\}
+\mpsi-{\frac12}\psiani+ (z)\right)\\
&=&z^{-1}\,\bracurr-1|\,\{1+\mpsi-{\frac12}\psiani+ (z)\},
\end{eqnarray*}
we have
\beqn
\bracurr-0|\,c(z)\opevol(t^\pm)
&=&\bracurr-1|\,e^{\phiani+(z)}\,\opevol(t^\pm),\nonumber\\
\bracurr-0|\,\gamma(z)\opevol(t^\pm)
&=&z\,\bracurr-1|\,\psiani- (z)e^{\phiani+(z)}\,\opevol(t^\pm),
\la{instC}\\
\bracurr-2|\,b(z)\opevol(t^\pm)
&=&z^{-1}\,\bracurr-1|\,\mpsi-{\frac12}\,e^{-\phiani+(z)}\,\opevol(t^\pm),
\nonumber\\
\bracurr-2|\,\beta(z)\opevol(t^\pm)
&=&\bracurr-1|\,\{1+\mpsi-{\frac12}\psiani+ (z)\}
\,e^{-\phiani+(z)}\,\opevol(t^\pm),
\la{instB}
\eeqn
where
\beq
\psiani+ (z)=\sum_{n\geq0}\ppsip-n\,z^{-n-1}
\quad\hbox{and}\quad
\psiani- (z)=\sum_{n\geq0}\mpsip-n\,z^{-n-2}.
\eeq
Next task is to see the effect of \(e^{\pm\phiani+(z)}\) on
\(\opevol(t^\pm)\). Here we note that for a formal power series
\(\tau(w)=\sum_{n\geq0}\tau_{n+\frac12}\,w^n \) with odd Grassmann
parity,
\beqn
\lefteqn{
e^{\phiani+(z)}\psi^+[\tau(\bullet)]e^{-\phiani+(z)}=
\oint_{|w|<|z|}dw\,\tau (w)(1-\frac{w}{z})\psi^+(w)}\nonumber\\
&=&\psi^+[\trsub[+,z]\tau(\bullet)],\quad\hbox{with}\quad
\trsub[+,z]\tau(w)\equiv (1-\frac{w}{z})\tau(w)
=\sum_{n=0}(\tau_{n+\frac12}-\frac1z\tau_{n-\frac12})\,w^n,\nonumber\\
\lefteqn{
e^{-\phiani+(z)}\psi^+[\tau(\bullet)]e^{\phiani+(z)}=
\oint_{|w|<|z|}dw\,\tau(w)\,\frac{1}{1-\frac{w}{z}}\,\psi^+(w)}\nonumber\\
&=&\psi^+[\trsub[-,z]\tau(\bullet)],\quad\hbox{with}\quad
\trsub[-,z]\tau(w)\equiv\frac{1}{1-\frac{w}{z}}\,\tau(w)
=\sum_{n=0}(\sum_{m=0}^n\tau_{n-m+\frac12}\,z^{-m})\,w^n,
\la{transf1}
\eeqn
and similarly,
\beqn
e^{\phiani+(z)}\psi^-[\tau(\bullet)]e^{-\phiani+(z)}&=&
\oint_{|w|<|z|}dw\,w\tau(w)\,\frac{1}{1-\frac{w}{z}}\,\psi^-(w)=
\psi^-[\trsub[-,z]\tau(\bullet)],\nonumber\\
e^{-\phiani+(z)}\psi^-[\tau(\bullet)]e^{\phiani+(z)}&=&
\oint_{|w|<|z|}dw\,w\tau(w)(1-\frac{w}{z})\psi^-(w)=
\psi^-[\trsub[+,z]\tau(\bullet)].\la{transf2}
\eeqn
Moreover we observe for a formal power series
\(x(w)=\sum_{n=1}x_n\,w^n\),
\beq
e^{\pm\phiani+(z)}e^{j^+[x(\bullet)]}
=e^{j^+[\trsub[\pm,z] x(\bullet)]},
\eeq
where
\beq
j^+[\trsub[\pm,z] x(\bullet)]=\sum_{n=1}\pj-n\,(x_n\mp\frac{z^{-n}}{n}),
\eeq
and hence
\beq
\trsub[\pm,z] x(w)\equiv x(w)\pm\log (1-\frac{w}{z}).
\eeq
Suppose that \(f(\tau(\bullet))\) be a functional of \(\tau(z)\)
and imagine \(\opT[z,\tau]\) be a differential
operator with respect to \(\{\tau_{n+\frac12}\}\) such that
\beq
e^{\opT[z,\tau]}\;f(\tau(\bullet))=f(\trsub[+,z]\tau(\bullet)).
\eeq
Clearly, the inverse operation gives
\beq
e^{-\opT[z,\tau]}\;f(\tau(\bullet))
=f(\trsub[-,z]\tau(\bullet)).
\eeq
One can see such an operator is given by
\beq
\opT[z,\tau]
=\sum_{n=0}\sum_{m=1}\frac{-1}{m}\,z^{-m}\;
\tau_{n+\frac12}\der-{\tau_{n+m+\frac12}}.
\eeq
Therefore we have
\beqn
e^{\pm\phiani+(z)}\,\opevol(t^\pm(\bullet))
&=&\opevol(\,\trsub[\pm,z]\pwptd-{\bullet},\;
\trsub[\mp,z]\pwmtd-{\bullet},\;
\trsub[\pm,z]\pwpte-{\bullet},\;\pwmte-{\bullet}\,)
\nonumber\\
&=&e^{\pm\{\,\pwlog-z\,\der-{\pt-{2n}}+
\opT[z,t^+_o]-\opT[z,t^-_o]\,\}}\;
\opevol(t^\pm(\bullet))\nonumber\\
&=&e^{\pm\pwlog-z\,\opd[+,2n]}\;\opevol(t^\pm(\bullet))
\la{expcr}
\eeqn
where
\(\opd[+,2n]\) is the operator defined previously in (\ref{oppmde}).
This is expected because from
\beq
\opd[\pm,2n+1]\,\opevol(t^\pm)=\pm\ppsip-n\;\opevol(t^\pm)
\la{diff-oppsi}
\eeq
and the fact that \(\opd[\pm,n],\>\;n=1,2,\cdots\) obeys
the same algebra of the annihilation part of \(\aljpsi\)
when \([\>,\>]_\pm\) is replaced by \(-[\>,\>]_\pm\),
with the correspondence
\[
\opd[\pm,2n+1]\leftrightarrow \pm\pmpsip-n,\qquad
\opd[\pm,2n]\leftrightarrow \pmj-n.
\]
\(\opd[\pm,2n]\) must give the same effect on
\(\opevol(t^\pm)\) as \(\pmj-n\) does.
Actually, one can observe for example
\beqn
e^{\pm\pwslog-z\,\opd[+,2n]}\opd[\pm,2m+1]e^{\mp\pwslog-z\,\opd[+,2n]}
&=&\sum_{n=0}z^{-n}\opd[\pm,2(m+n)+1],\nonumber\\
e^{\pm\pwslog-z\,\opd[+,2n]}\opd[\mp,2m+1]e^{\mp\pwslog-z\,\opd[+,2n]}
&=&\opd[\mp,2m+1]-\frac1w\,\opd[\mp,2(m+n)+1],
\eeqn
which are equivalent to (\ref{transf1}) and (\ref{transf2})
when making replacement
\(\opd[\pm,2n+1]\rightarrow \pm\pmpsip-n\) and
\(-\opd[\pm,2n]\rightarrow \pmj-n\).
Now from (\ref{instC}), (\ref{instB}), (\ref{expcr}) and
(\ref{diff-oppsi}), we obtain
\beqn
v_c\zpmt&=&e^{\pwslog-z\,\opd[+,2n]},\nonumber\\
v_\gamma\zpmt&=&e^{\pwslog-z\,\opd[+,2n]}
\sum_{m=0}-z^{-(m+1)}\opd[-,2m+1]\nonumber\nonumber\\
&=&-z^{-1}\,\opd[-,1]\,e^{\pwslog-z\,\opd[+,2n]},\nonumber\\
v_b\zpmt&=&e^{-\pwslog-z\,\opd[+,2n]}\,(-z^{-1}\,\opd[-,1]),\nonumber\\
v_\beta\zpmt&=&e^{-\pwslog-z\,\opd[+,2n]}\,
(\;1+\sum_{m=0}z^{-(m+1)}\opd[+,2m+1]\opd[-,1]\;)\nonumber\\
&=&e^{-\pwslog-z\,\opd[+,2n]}+ z^{-1}\,\opd[+,1]\,
e^{-\pwslog-z\,\opd[+,2n]}\,\opd[-,1].
\eeqn
Moreover we can find the expressions for \(x\not=0\) and \(\xi\not=0\),
by noting
\beqn
\lefteqn{
v_\alpha\zxxipmt\bracurr-1|\,e^{\opxxi}\opevol(t^\pm)}\nonumber\\
&=&e^{\xi(\opd[+,1]-\opd[-,1])}\,v_\alpha\zpmt\,
e^{-\xi(\opd[+,1]-\opd[-,1])}\;\bracurr-1|\,e^{\opxxi}\opevol(t^\pm)
\eeqn
where the subscript \(\alpha\) stands for \(c\), \(\gamma\),
\(b\) and \(\beta\). Using the above formula and taking into
account\footnote{ The fact that
\[
e^{\xi(\opd[+,1]-\opd[-,1])}e^{-\xi D}=
e^{ \xi\sum{\bar t}_{2m+1}\der-{t_{2(m+1)}}}\;
e^{\xi(\der-{\pt-1} -\der-{\mt-1}-\der-\xi)}
\]
acts as an identity on \(e^{\opxxi}\opevol(t^\pm)\) implies
the relation (\ref{opxxi-x}).}
\beq
D\,e^{\opxxi}=\sigma\,e^{\opxxi}
\eeq
and
\[
\bracurr-1|\,\sigma=\bracurr-1|(\ppsi-{\frac12}+\mpsi-{\frac12})
=\bracurr-1|\mpsi-{\frac12},
\]
we obtain consequently
\beqn
v_c\zxxipmt
&=&
e^{\pwslog-z\,\{\;\opd[+,2n]+\xi(\opd[+,2n+1]+\opd[-,2n+1])\;\}},
\nonumber\\
v_\gamma\zxxipmt
&=&
z^{-1}D
e^{\pwslog-z\,\{\;\opd[+,2n]+\xi(\opd[+,2n+1]+\opd[-,2n+1])\;\}},
\nonumber\\
v_b\zxxipmt
&=&
z^{-1}e^{-\pwslog-z\,\{\;\opd[+,2n]+\xi(\opd[+,2n+1]+\opd[-,2n+1])\;\}}
\,D,
\nonumber\\
v_\beta\zxxipmt
&=&
e^{-\pwslog-z\,\{\;\opd[+,2n]+\xi(\opd[+,2n+1]+\opd[-,2n+1])\;\}}
\nonumber\\
&&\quad
-z^{-1}\{\opd[+,1]+\xi\der-x\}\,
e^{-\pwslog-z\,\{\;\opd[+,2n]+\xi(\opd[+,2n+1]+\opd[-,2n+1])\;\}}
\,D.
\eeqn
Let us reflect on the results from the view point of
the superbosonization.
We have first realized the negative frequency part of \(\gl11\)
as differential operators which act on the
polynomial ring \({\cal B}=\complnum [[\pt-m,\,\mt-m;\>m\geq 1\,]]\).
Then, based on the Kac-van de Leur superbosonization, we
have obtained the description of the B-C system on \({\cal B}\).
(More precisely, we must introduce one more variable which counts the
total charge.)
Consider the vertex operators
\beqn
V_c\zpmt\,\bracurr-1|\,\opevol(t^\pm)=
\bracurr-0|\,\opevol(t^\pm)\,c(z),&&\;
V_b\zpmt\,\bracurr-1|\,\opevol(t^\pm)=
\bracurr-2|\,\opevol(t^\pm)\,b(z),\nonumber\\
V_\gamma\zpmt\,\bracurr-1|\,\opevol(t^\pm)=
\bracurr-0|\,\opevol(t^\pm)\,\gamma (z),&&\;
V_\beta\zpmt\,\bracurr-1|\,\opevol(t^\pm)=
\bracurr-2|\,\opevol(t^\pm)\,\beta (z).
\eeqn
{}From (\ref{vopC}) and (\ref{vopB}) and the expressions of
\(A_M\) and \({\widetilde A_M}\), these are given by
\beqn
V_c\zpmt&=&
e^{\pwpte-z -\frac12 zt^+_o\pwmtd-z}\;
\{\,v_c\zpmt+z\pwmtd-z\,v_\gamma\zpmt\,\},\nonumber\\
V_\gamma\zpmt&=&
e^{\pwmte-z +\frac12 zt^+_o\pwmtd-z}\;
\{\,v_\gamma\zpmt+\pwptd-z\,v_c\zpmt\,\},\nonumber\\
V_b\zpmt&=&
e^{-\pwpte-z -\frac12 zt^+_o\pwmtd-z}\;
\{\,v_b\zpmt+\pwptd-z\,v_\beta\zpmt\,\},\nonumber\\
V_\beta\zpmt&=&
e^{-\pwmte-z +\frac12 zt^+_o\pwmtd-z}\;
\{\,v_\beta\zpmt -z\pwmtd-z\,v_b\zpmt\,\},
\eeqn
which may be viewed as a kind of superbosonization
realized on \({\cal B}\).
The expression of the bilinear identity for the \(\tau\)-function
(super-Hirota equation) can be obtained without any difficulty
although we have not written down here.

\resection{Concluding Remarks}
We have established the operator theory for the SKP hierarchies and
formulated the maximal SKP hierarchy which includes all the flows of
the known SKP hierarchies in a unified way and allows the
\(\tau\)-function description of the theory. Now the relation
between different SKP hierarchies is very clear. \\ \indent
Here let us look briefly at the geometrical aspects of the subject,
which we have not touched on so far. As stated in Ref.\cite{mula}, for
a general B-C system, the fields B and C are defined on an arbitrary
\((1|1)\) dimensional complex supermanifold (i.e., a super curve in
general without any superconformal structure), and are considered as
sections of line bundles \(\omega^k\) and
\(\omega^{1-k}\) (in our setting,
\(k=1\)) respectively. According to the coordinate transformation
\((z,\,\theta)\rightarrow ({\tilde z},\,{\tilde \theta})\),
a section \(\sigma\in \omega\) transforms as
\(\sigma={\tilde \sigma}
\frac{\partial ({\tilde z},\,{\tilde \theta})}{\partial (z,\,\theta)}\).
It should be noted here that the moduli space of supercurves is
coincident with that of the \(N=2\) super-Riemann
surface\cite{schwarz}.  The hidden \(N=2\) superconformal symmetry of
the B-C system originates in this coincidence.  In Ref.\cite{mula-rab}
the super-Krichever construction was studied and the geometrical
meaning of the SKP hierarchies was clarified. Their arguments ought
to be traced in the field theoretic context. The super-Krichever map
assigns injectively a point of the {\sG}, hence a state in the B-C
Fock space, to a set of geometrical data of an arbitrary supercurve
and a line bundle
\(\omega^k\) on it. The same map was also investigated in constructing
the operator formalism for the superstring
theory\cite{alvarez-g}. From these studies it becomes clear that as
long as we work within the frame of our restricted {\sG} \(\G\) or
equivalently the space of the wave operators \(\sgrass\), the
supercurves we can cover are constrained excluding super-Riemann
surfaces of genus \(g\neq 1\). This is a disappointing fact. However
if we convert the description of the theory in terms of the
\(\tau\)-function, we would be able to generalize the theory to cover
more general {\sG s}. In this relation, comparison with the
\(\tau\)-function of A.S. Schwarz\cite{schwarz2} which is defined
in more abstract and general manner would be helpful. An
interesting problem is to find, at least in the genus one case, some
characterization of the locus of the geometrical data coming from
\(N=1\) super-Riemann surfaces inside the larger moduli space of the
geometrical data coming from general supercurves.  To think over
these issues, it would be useful to recall another SKP hierarchy given
by LeClair\cite{leclair}.  It is based on the superbosonization
which preserves the superconformal symmetry
manifestly\cite{taka-MS}. In this superbosonization the B-C system is
expressed in terms of the following currents:
\begin{eqnarray*}
&&J=-BC,\qquad J^\ast=-D\log B,\\
&&J(z_1,\theta_1)J^\ast (z_2,\theta_2)
\sim\frac{1}{z_1-z_2-\theta_1\theta_2},\quad
J(1)J(2)\sim{\rm regular},\quad
J^\ast (1)J^\ast (2)\sim{\rm regular}.
\end{eqnarray*}
The \(\tau\)-function of this SKP hierarchy is defined as
\[
\tau(t,\,t^\ast)=\bracurr-1|\,
e^{J^\ast[t^\ast(\bullet)]+J[t(\bullet)]}\,
|\ketcurr-{{\Xi_0}_{\ltinf}},
\]
where \(J[t(\bullet)]\) stands for the expression (\ref{hamil-jskp}) and
\(J^\ast[t^\ast(\bullet)]\) is defined similarly.
{}From the above expression we recognize that this SKP hierarchy is
essentially JSKP hierarchy, since the former reduces to the latter by
setting \(t_n^\ast=0,\>\;n=1,2,\cdots\). On the other hand we can see
that the state
\(e^{J^\ast[t^\ast(\bullet)]}|\ketcurr-{{\Xi_0}_{\ltinf}}\) does not
satisfy the bilinear identity. The operator
\(e^{J^\ast[t^\ast(\bullet)]}\) cannot be considered as the time
evolution operator on the {\sG} and is introduced in order to keep the
whole information of the state \(|\ketcurr-{{\Xi_0}_{\ltinf}}\).
{}From the speciality of this superbosonization, when the initial state
\(|\ketcurr-{{\Xi_0}_{\ltinf}}\) is associated with a
super-Riemann surface, its superconformal structure is preserved
and translated into the one expressed in terms of the time variables
\(t_n\) and \(t_n^\ast\).\\
In addition to the geometrical consideration, there are several
directions for future work.  For the KP hierarchy N soliton
solutions are obtained by successively applying the vertex
operators to the vacuum state. The same procedure would be also
applicable for the SKP hierarchies.  The
operator theory for the SBKP hierarchy and its
\(\tau\)-function description has already been studied in
\cite{kac-vdleur2} (see also \cite{ike-yama}).
Making similar efforts as we did in the previous section, we will be
able to transfer their elaborated result to the SBKP version of our
maximal SKP hierarchy, which will be more accessible.  Though it is
not clear at present that the SKP hierarchies have relevance to
physics such as the two-dimensional supergravity, it would be
interesting on its own to seek additional symmetries, the relation to
super-\(W\) algebras and favorable reductions to the KdV type together
with their
\(\tau\)-function description.


\vskip 8mm\noindent
{\large \bf Acknowledgement}\\\noindent The auther is very grateful to
Prof. Ryu Sasaki for continuous encouragement during the work and
careful reading of the manuscript.
\vskip 8mm\noindent

\appendix
\resection{Calculation of \(\bracurr-2|\,B_j\,|X_{\ltinf}\rangle\)}
\la{append; a}
Let us show
\beq
\frac{1}{\bracurr-1|X_{\ltinf}\rangle}
\bracurr-2|\,B_j\,|X_{\ltinf}\rangle
=\frac{(-1)^{j+1}}{{\rm sdet}_{\pi}X_{\wltinf}}\;
{\rm sdet}_{\pi}X^{(j)}_{\wltinf},\qquad j=-1,0,1\cdots.
\la{correB}
\eeq
Using the formula (\ref{correl}) we have
\beq
\bracurr-2|\,B_j\,|X_{\ltinf}\rangle=
\int\prod_{r>0}d\tilgam-r
\int\prod_{l\geq 1}d\tilc-l\;
\tilgam-{\frac12}(-1)^j\der-{\tilcapc-{-j}}\left.
\prod_{k\geq 1}\tilc-{}(X)_k\prod_{s>0}\delta(\tilgam-{}(X)_s)
\right|_{\tilcapc-{\ltinf}=0}.
\eeq
Then we change the integration variables
\(\tilcapc-j\rightarrow\tilcapc-j'=\tilcapc-{}(X)_j\)
as in (\ref{ch-var}), provided that in this case \(j=1,2,\cdots\) and
\(\tilcapc-{\leinf}\) are considered as auxiliary variables.
The right-hand side of the above expression becomes
\beqn
\lefteqn{
(-1)^j\;{\rm sdet}X_{\wltinf}\;
\int\prod_{r>0}d\tilgam-r'\int\prod_{l\geq 1}d\tilc-l'}\nonumber\\
&&\sum_{k>0}(\,\tilcapc-k' - \sum_{l\geq 0}\tilcapc-{-l}X_{l,-k}\,)
({X_{\wltinf}}^{-1})_{-k,-1}\,
\sum_{m>0}X_{j,-m}\der-{\tilcapc-{m}'}
\left.\prod_{k\geq 1}\tilc-k'\prod_{s>0}\delta(\tilgam-s')
\right|_{\tilcapc-{\ltinf}=0}.
\eeqn
Integrating by part we have
\beq
=(-1)^{j+1}\,{\rm sdet}X_{\wltinf}\cdot
\sum_{k>0}X_{j,-k}({X_{\wltinf}}^{-1})_{-k,-1}.
\eeq
Since \(\bracurr-1|X_{\ltinf}\rangle={\rm sdet}X_{\wltinf}\), we thus obtain
\beq
\frac{1}{\bracurr-1|X_{\ltinf}\rangle}
\bracurr-2|\,B_j\,|X_{\ltinf}\rangle
=(-1)^{j+1}\,\sum_{k>0}X_{j,-k}({X_{\wltinf}}^{-1})_{-k,-1}.
\eeq
It is not hard to see that
\beq
\frac{1}{{\rm sdet}_{\pi}X_{\wltinf}}\;
{\rm sdet}_{\pi}X^{(j)}_{\wltinf}=
\sum_{k>0}X_{j,-k}({X_{\wltinf}}^{-1})_{-k,-1},\la{co-fac}
\eeq
which is viewed as a co-factor expansion of the superdeterminant
\({\rm sdet}_{\pi}X^{(j)}_{\wltinf}\).
Taking into account
\beq
({X_{\wltinf}}^{-1})^{01}=
-{X_{\wltinf}^{00}}^{-1}X_{\wltinf}^{01}
{\widehat X_{\wltinf}^{11}}{}^{-1}\quad\hbox{and}\quad
({X_{\wltinf}}^{-1})^{11}=
{\widehat X_{\wltinf}^{11}}{}^{-1}
\eeq
where the notation (\ref{hatmat}) is used, we can reduce
(\ref{co-fac}) to an exercise for an ordinary matrix determinant.

\resection{More about the superbosonization}
\la{append; b}
As mentioned in \(\S\)\ref{sec:tau}, in the \(\beta\)-\(\gamma\)
system, local fields which carry the picture charge exist
and are most easily identified in the
FMS bosonization\cite{fms}.
These are
\beq
\delta (\beta(z))=
e^{\mphi-0}z^{-\mj-0}e^{\phicri- (z)}e^{\phiani- (z)},\qquad
\delta (\gamma(z))=
e^{-\mphi-0}z^{\mj-0}e^{-\phicri- (z)}e^{-\phiani- (z)},
\la{delta}\eeq
and the additional free fermionic pair independent of
\(\delta (\beta(z))\) and \(\delta (\gamma(z))\),
\beq
\xi (z)=\Theta (\beta(z))\quad\hbox{and}\quad
\eta (z)=\der-z\Theta (\gamma(z)),\la{xieta}
\eeq
where \(\Theta\) is Heviside step function\cite{verlin**2}.
FMS bosonization says the original \(\beta\), \(\gamma\) fields
are reconstructed from the above fields as
\beq
\beta(z)=\der-z\Theta (\beta(z))\cdot\delta (\gamma(z)),\qquad
\gamma(z)=\der-z\Theta (\gamma(z))\cdot\delta (\beta(z)).
\la{verlin}
\eeq
Obviously the formulas (\ref{delta}) are considered to be the
same as in the superbosonization we discussed. With this in mind
we can find from (\ref{xieta}) or (\ref{verlin})
the superbosonization rules
for \(\xi\) and \(\eta\) fields:
\beqn
\eta(z)&=&\mathop{\lim}_{w\rightarrow z}\>
\psi^-(z)\,c(w)\delta (\gamma(w))\nonumber\\
&=&e^{\pphi-0-\mphi-0}e^{\phicri+ (z)-\phicri- (z)}\psi^-(z)\,
z^{\pj-0+\mj-0}\,e^{\phiani+ (z)-\phiani- (z)},\nonumber\\
\der-z\xi(z)&=&\mathop{\lim}_{w\rightarrow z}\>
\psi^+(z)\,b(w)\delta (\beta(w))\nonumber\\
&=&e^{-\pphi-0+\mphi-0}e^{-\phicri+ (z)+\phicri- (z)}\psi^+(z)\,
z^{-(\pj-0+\mj-0)}\,e^{-\phiani+ (z)+\phiani- (z)}.
\eeqn
These expressions already appeared in \cite{bergv2}


\end{document}